\documentclass[12pt,preprint]{aastex}

\slugcomment{\today}

\shorttitle{The Most Likely Sources of Cosmic-Ray Electrons}
\shortauthors{Kobayashi et al.}

\begin{document}


\title{The Most Likely Sources of High Energy Cosmic-Ray Electrons \\
in Supernova Remnants}


\author{T.Kobayashi}
\affil{Department of Physics, Aoyama Gakuin University, 
Sagamihara 229-8558, Japan}

\author{Y.Komori}
\affil{Kanagawa University of Human Services, Yokosuka 238-0013, Japan}

\author{K.Yoshida}
\affil{Faculty of Engineering, Kanagawa University, Yokohama 221-8686, Japan}
\email{yoshida@kit.ie.kanagawa-u.ac.jp}

\and

\author{J.Nishimura}
\affil{The Institute of Space and Astronautical Science, 
Sagamihara 229-8510, Japan}




\begin{abstract}
Evidences of non-thermal X-ray emission and TeV gamma-rays from 
the supernova remnants (SNRs) has strengthened the hypothesis 
that primary Galactic cosmic-ray electrons are accelerated in SNRs. 
High energy electrons lose energy via synchrotron and inverse Compton
processes during propagation in the Galaxy. 
Due to these radiative losses, 
TeV electrons liberated from SNRs at distances 
larger than $\sim$1 kpc, or times older than $\sim10^5$ yr, 
cannot reach the solar system. 
We investigated the cosmic-ray electron spectrum observed in the solar system 
using an analytical method, and 
considered several candidate sources among nearby SNRs 
which may contribute to the high energy electron flux. 
Especially, we discuss 
the effects for the release time from SNRs after the explosion, 
as well as the deviation of a source spectrum from a simple power-law. 
From this calculation, 
we found that some nearby sources such as the Vela, Cygnus Loop, or Monogem 
could leave unique signatures 
in the form of identifiable structure 
in the energy spectrum of TeV electrons and 
show anisotropies towards the sources, 
depending on when the electrons are liberated from the remnant. 
This suggests that, 
in addition to providing information on the mechanisms of acceleration 
and propagation of cosmic-rays, 
specific cosmic-ray sources can be identified through 
the precise electron observation in the TeV region. 
\end{abstract}




\section{Introduction}
\label{sec:intro}


Radio observations have indicated that supernova remnants (SNRs) are 
the most likely sources of cosmic-ray electrons in the energy region 
below $\sim10$~GeV. 
Evidence for non-thermal X-ray emission 
from the supernova remnant SN1006 discovered 
with the ASCA satellite \citep{koyama95} 
strongly supports the hypothesis that
Galactic cosmic-ray electrons in the TeV region originate in supernovae.  
In this case, 
TeV gamma-rays should be produced via the inverse Compton process 
between accelerated electrons and the cosmic microwave 
background (CMB) radiation, 
and indeed TeV gamma-rays were detected by the CANGAROO 
experiment \citep{tanimori98}.
Additional
evidence for X-ray synchrotron emission is provided by
observations of several other SNRs, such as 
RX J1713.7-3946 \citep{koyama97}, 
Cas A \citep{allen97}, 
IC443 \citep{keohane97}, 
G374.3-0.5 \citep{slane99}, 
and RX J0852.0-4622 \citep{slane01}. 
In SNR RX J1713.7-3946, 
evidence for the acceleration of cosmic-ray protons is 
reported by \citet{enomoto02} from TeV gamma-ray observations. 
However, there are also arguments that the TeV gamma-rays 
observed cannot be interpreted as hadronic in origin 
\citep{reimer02, butt02}.


To clarify the origins of cosmic rays 
and their propagation mechanisms within the Galaxy, 
electrons provide an ideal probe, complementary to the nucleonic component, 
due to their low mass and leptonic nature. 
High energy electrons lose energy primarily via synchrotron 
and inverse Compton processes during propagation in the Galaxy. 
These processes, combined with the absence of hadronic interactions, 
simplify modeling of the propagation of electrons compared with other
cosmic-ray components such as nucleons. 
Since the energy loss rate is almost proportional to $E^2$, 
higher energy electrons lose energy more rapidly. 
TeV electrons lose most of their energy on a time scale of $\sim10^5$yr, and 
their propagation distances are therefore limited to several hundred pc.

Measurements of cosmic-ray electrons in the TeV region 
have been successfully performed only with emulsion chamber detectors
\citep{nishimura80, kobayashi99}. 
The observed energy spectra extend without cut-off up to $\sim2$TeV. 
This means that 
the observed TeV electrons must have been 
accelerated in SNRs at distances 
within several hundred pc, and 
at times within $\sim10^5$ yr ago.


%
%
Several models have been proposed to describe cosmic-ray propagation 
in the Galaxy, 
motivated by attempts to fit existing data on the heavy primary 
composition and spectra 
of different species assuming the same simple power-law source
spectrum. 
Among these models,  
it is believed that the diffusion model provides the most realistic 
description of the propagation. 
The solution for the electron density in the diffusion equation 
is derived by several authors (e.g. \citealt{jokipii68, ginzburg76,
nishimura79, berezinskii90} and other references therein), 
under different boundary conditions for the Galactic halo.

In a study of the propagation of cosmic-ray electrons, 
\citet{shen70} first pointed out that a continuous source 
distribution model is not valid in the energy region above 100GeV, 
because the electron spectrum in that energy range 
depends upon the age and distance of a few local sources. 
He showed that high energy electrons over 100GeV probably 
come from the Vela X pulsar, 
and predicted a cutoff at $\sim$2 TeV in the energy spectrum. 
Although various parameters for SNRs and diffusion coefficients 
were not known very well at that time, 
the concepts proposed by the author 
has been accepted in later work.

\citet{cowsik79} have investigated 
the electron spectrum in terms of 
contributions from discrete sources such as SNRs 
distributed over the Galaxy. 
Their calculations apply in the 
the case of sources continuously active in time. 
They suggested that such sources must be located within a few hundred
pc of the solar system to prevent
radiative energy losses from inducing a premature 
cut-off in the energy spectrum. 
They took the diffusion coefficient $D=10^{28}$ cm$^{2}$s$^{-1}$, 
independent of energy, which is 
much smaller than recently accepted values 
in the energy region over 100~GeV 
(see section \ref{subsec:diffusion}). 
They concluded that it is very unlikely that SNRs are  
the only sources of cosmic-ray electrons in the energy range
1GeV $-$ 1TeV, but this result
seems due to the inappropriate value of  
the diffusion coefficient assumed by themselves.

\citet{nishimura79} calculated the propagation of electrons from SNRs, 
using a solution of the diffusion equation 
with a disk-shaped boundary condition. 
They showed that the TeV electron spectrum would 
deviate from power-law behavior 
due to fluctuations caused by the small number of sources 
capable of contributing to the observed flux in the TeV region.

To explain the features of both energy spectrum 
and charge composition of electrons, 
\citet{aharonian95} suggested an approach that separates 
the contribution of one nearby source, with age $10^5$yr 
and distance 100pc, 
from the contributions of other Galactic sources at distances 
over 1kpc. 
Using this approach, 
\citet{atoyan95} calculated the electron energy spectrum
in analytical form. 
They considered the energy-dependent diffusive propagation of electrons, 
assuming the diffusion coefficient 
$D=7\times10^{27}$cm$^2$s$^{-1}$ at 10 GeV, 
increasing with energy as $E^{0.6}$. 
They showed that the observed energy spectrum 
from sub-GeV to TeV energies can be explained by this model.

\citet{pohl98} has pointed out 
that, if cosmic-ray electrons are accelerated in SNRs, 
the observed electron spectrum should be strongly time-dependent 
above 30 GeV,  
due to Poisson fluctuations 
in the number of SNRs within a given volume and 
time interval. 
They assumed the diffusion coefficient 
$D=4{\times}10^{27}(E/1{\rm GeV})^{0.6}$cm$^2$s$^{-1}$, and 
calculated 400 cases of spectra using a random distribution of SNRs 
in space and time. 
They argued that in their model, 
an injection spectral index of 2.0 cannot be rejected 
for the observed local electron flux at high energy,
with a probability of 5\% over all their calculated spectra.

\citet{erlykin02} considered the model for 
cosmic-ray electron production in the supernova explosion 
and variants for the subsequent propagation. 
In their model, electrons are accelerated throughout the SNR life time 
of several $10^4$~yr, and after the cease of the acceleration process 
electrons leave the remnant. 
They simulated the electron spectra 
by taking the random distribution of SN in the Galaxy and 
letting the electrons diffuse to the solar system.

%
As pointed out in these works, 
at energy higher than $\sim$1 TeV, 
the propagation lifetime of electrons is so short that 
only a few cosmic-ray sources can contribute to the observed flux. 
Thus one expects fluctuations in the local electron energy spectrum 
reflecting the contributions of these small number of discrete sources. 
To analyze fluctuations in the spectrum, 
we use the method of separating the contributions 
of distant and nearby sources to the total flux of high energy electrons. 
Although this approach was discussed by earlier authors 
(e.g. \citealt{atoyan95} and references therein),  
the main difference between the present and earlier works is that in this paper
we took only known, observed local SNRs as nearby sources. 
Using information on the ages and distances of observed SNRs 
in the neighborhood of the solar system, 
we could determine which sources contribute electrons 
most efficiently in a given energy range, and thus 
correlate individual sources with features of the electron spectrum. 
In addition, one would expect the high energy electron flux to be anisotropic 
if the most significant sources are nearby 
\citep{shen71, ptuskin95}. 
This means that 
we can identify cosmic-ray electron sources from 
the analysis of the local electron spectrum 
together with anisotropies in arrival direction.

For this purpose, 
we calculated the electron energy spectra 
with the diffusion model in a semi-analytic approach, 
separating the contributions of distant and nearby sources. 
In this calculation, we particularly 
took into account of the effects of the delay 
of electron liberation from SNRs after the explosion, 
as well as the deviation from a single power-law spectrum of electrons 
in SNRs. 
As a result, we indicate that 
nearby SNRs such as the Vela, Cygnus Loop, or Monogem 
are the most likely candidates for 
the sources of cosmic-ray electrons in TeV region, 
depending on when the electrons are released from SNRs. 
If these sources are proven to be the main contributors 
of high energy electrons, 
we can precisely analyze the propagation of TeV electrons 
by comparing predicted results with observed electron spectra, 
and the degree of anisotropy in the direction of the sources. 
Such analyses will yield more detailed information about the acceleration 
and propagation of cosmic rays.


%
\section{Acceleration in SNRs and Propagation in the Galaxy}
\label{sec:propagation}

\subsection{Accelerated Electrons in SNRs and their Liberation Time}
\label{subsec:acceleration}

According to shock-acceleration models, 
the maximum energy of accelerated electrons is limited, 
by the SNR age, a free escape from the shock region of MHD turbulence, 
or synchrotron losses \citep{sturner97, reynolds96}. 
Analysis of the observed radio and X-ray spectra also indicates that 
the typical electron spectrum produced in a remnant 
is a power-law with a cut-off of ${\sim}10-100$TeV 
\citep{reynolds99, hendrick01}. 
Therefore, we approximated an electron injection spectrum of a power-law 
with an exponential cut-off of the form of ${\exp}(-E/E_{\rm c})$, 
as taken by these authors.

As discussed in \citet{erlykin02}, 
it is also important to take account of 
when the accelerated electrons are liberated from the remnant 
after the explosion. 
From shock-acceleration theory, it is suggested that 
the electrons are liberated from SNRs in the termination of the shock, i.e. 
when the shock velocity has dropped to the mean ISM Alf{\'{v}}en
velocity and the remnant has been reduced in the ISM 
(e.g. \citealt{dorfi00}). 
This time scale is $\sim10^5$~yr. 
However, there are no clear observational evidences 
for the release time of the TeV electrons from SNRs. 
In the scenario of the SNR origin of cosmic-ray electrons, 
the accelerated TeV electrons should be liberated from the remnant 
within ${\sim}10^{5}$yr, 
or the acceleration process should operate till the last phase of SNRs. 
This is because the life time of TeV electrons in the remnant 
is estimated to be less than a few times of $10^{4}$yr 
due to synchrotron loss inside SNRs with a magnetic field of 10${\mu}$G. 
Therefore, we evaluated the cosmic-ray electron spectra, 
changing the electron release time $\tau$ from 0 to $1\times10^5$~yr.

%
%
\subsection{Properties of the Propagation of High Energy Electrons}
\label{subsec:diffusion}

%
%
High energy electrons above 10 GeV lose their energy 
mainly via the synchrotron and inverse Compton processes 
while propagating through the Galaxy. The energy loss rate is given by 
\begin{equation}
 -\frac{dE}{dt}=bE^2,
\end{equation}
with
\begin{equation}
 b=\frac{4{\sigma}c}{3(mc^2)^2}(\frac{B^2}{8{\pi}}+{\rho_{ph}}). 
 \label{eq:b}
\end{equation}
Here, 
$E$ is the electron energy, 
$m$ is the mass of electron, 
$c$ is the speed of light, 
$B$ is the magnetic field strength in the Galaxy, 
$\rho_{ph}$ is the energy density of interstellar photons, 
and $\sigma$ is the cross-section for Thomson scattering. 
This formula (\ref{eq:b}) is valid when the inverse Compton process 
is well approximated by the Thomson scattering cross-section, 
but we need corrections from the Klein-Nishina formula 
for high energy electrons as described below.

Typically quoted values for the interstellar 
magnetic field are estimated from measurements of the Zeeman splitting, 
the Faraday rotation, or the radio synchrotron emission. 
The estimates from Zeeman splitting and Faraday rotation measurements 
refer only to the line of sight component of the magnetic field. 
They are also biased toward cold neutral and warm ionized regions, 
respectively. 
Using the radio synchrotron emission from relativistic electrons, 
the local magnetic field strength of $B_{\perp}{\simeq}5$~$\mu$G 
is obtained as a more adequate value \citep{ferriere01}. 
Here, $B_{\perp}$ means the magnetic field perpendicular to the electron
velocity, that is $B^{2}_{\perp}=2B^{2}/3$. 
In this calculation, we adopted $B_{\perp}=5 \mu$G for our Galaxy.

\begin{figure}
\epsscale{0.80}
 \plotone{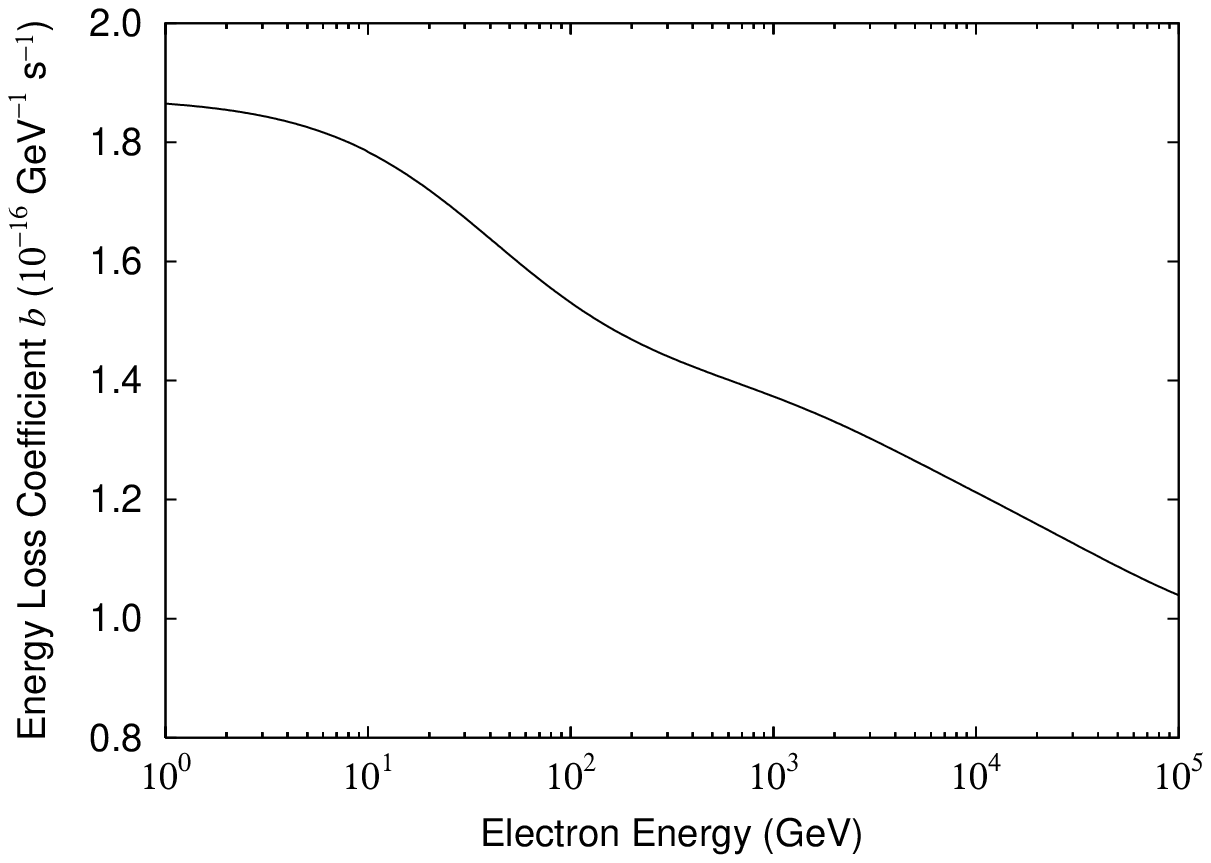}
 \caption{\label{fig:enelossb}
 Energy loss coefficient of cosmic-ray electrons in the Galaxy 
 with energy. 
 The magnetic field is assumed to be $B_{\perp}=5{\mu}$G. 
 }
\end{figure}

\begin{center}
\vspace*{0.5cm}
\framebox[3cm]{Figure \ref{fig:enelossb}}
\vspace*{0.5cm}
\end{center}

Since the Thomson scattering cross-section approximation 
is inadequate for high energy electrons, 
we evaluate the energy loss coefficient $b$ using 
the Compton scattering cross-section in the following. 
The interstellar radiation field in the Galaxy 
is dominated by three components: 
2.7K cosmic microwave background (CMB), 
re-emitted radiation from dust grains, and stellar radiation. 
The energy densities of photons are 
0.26~eV/cm$^3$ for CMB, 
0.20~eV/cm$^3$ for re-emitted radiation from dust grains, 
and 0.45~eV/cm$^3$ for stellar radiation, 
respectively \citep{mathis83}.
For inverse Compton scattering, 
we calculated the Klein-Nishina cross-section accurately 
using the formula by \citet{blumenthal70}. 
The resulting 
energy loss coefficient $b$ decreases gradually with energy 
as shown in Figure \ref{fig:enelossb}, 
where the radiation field is assumed to be isotropic. 
For electrons in the TeV region, 
energy loss through interactions with the CMB dominates over 
the other two components.

%
%
In a diffusive propagation model, 
the diffusion coefficient determines the travel distance 
of electrons in a given time. 
The diffusion coefficient $D$ 
has been estimated by using the observed ratios of 
secondary to primary nuclei (B/C) by 
HEAO-C \citep{engelmann90} and Voyager \citep{lukasiak94}, 
and is given by 
\begin{equation}
 D = 2\times10^{28}(E/5{\rm GeV})^{\delta} ({\rm cm^2/s}), 
  \label{eq:lowD}
\end{equation}
where 
\[ \delta = \left\{ 
 \begin{array}{@{\,}ll}
  0   & \mbox{($E<5${\rm GeV})}    \\
  0.6 & \mbox{($E{\geq}5${\rm GeV})}. \\
 \end{array}
 \right. \]
The formula (\ref{eq:lowD}) agrees well with the experimental data 
up to tens of GeV. 
Beyond 100GeV, there have been scarcely data on B/C until quite recently, 
and it was not clear whether the formula (\ref{eq:lowD}) 
is applicable to the higher energy region or not.

There are arguments that simple extrapolation of 
$D {\propto} E^{\delta}$ with ${\delta}=0.6$ 
gives too large diffusion coefficient in $1-100$~TeV region 
to interpret the observed anisotropy of primary cosmic-rays 
less than $10^{-3}$ (\citealt{ambrosio03} and other references therein). 
Thus, it has been argued that the increase of the diffusion coefficient 
with energy should become slower than  ${\delta}=0.6$ 
at somewhere beyond tens of GeV, and 
also that ${\delta}=0.3$ is plausible from 
a Kolmogorov-type spectrum of turbulence in the interstellar medium 
(e.g. \citealt{gaisser90, gaisser00, jones01}). 
In fact, 
a recent work of emulsion chamber experiments with long balloon exposures 
has shown the data of B/C and (secondary species of Fe)/Fe 
in TeV region, 
which indicate that the diffusion coefficient is consistent with 
the change of ${\delta}$ from 0.6 to 0.3 between tens of GeV and 1TeV 
\citep{furukawa03}. 

Hence we take 
$D = (2-5)\times10^{29}(E/{\rm TeV})^{0.3} ({\rm cm^2 s^{-1}})$ 
as a plausible diffusion coefficient in TeV region. 
The numerical values of 2 and 5 
are derived from the assumption 
that the formula (\ref{eq:lowD}) is valid up to 50~GeV 
or 1~TeV, respectively. 
With this assumption, 
the source spectral index $\gamma$ is chosen to maintain 
$\gamma + \delta = 2.7$.

%
%
Since high energy electrons lose energy by synchrotron and inverse Compton
processes at the rate $dE/dt = -bE^2$, 
electrons lose almost all of their energy $E$ 
after time $T=1/bE$. 
Therefore, electrons observed with energy $E$ 
must have been accelerated 
within $T = 1/bE$ from the present. 
Hence, the lifetime $T$ becomes progressively shorter with increasing
energy. 
Assuming $B_{\perp}=5 \mu$G and 
taking the Klein-Nishina formula for Compton process,
the lifetime is $T = 1/bE = 2.5\times10^5({\rm yr})/E({\rm TeV})$. 
As an approximate treatment, 
electrons can diffuse a distance of $R = (2DT)^{1/2}$ 
during this time; 
i.e. $0.6-0.9$kpc for $D=(2-5)\times10^{29} ({\rm cm^2 s^{-1}})$
at 1 TeV. 
More precise estimate for $R$ is given in the following section.

\subsection{Propagation of Cosmic Rays from a Single Source}

In the diffusion model for the propagation of electrons in the
Galaxy, 
the electron density $N_{\rm e}$ is given by the equation 
\begin{equation}
\frac{d{N_{\rm e}}}{dt} - {\nabla}(D{\nabla}N_{\rm e}) 
- \frac{\partial}{{\partial}E}(bE^{2}N_{\rm e}) = Q(E,r,z,t), 
\label{eq:diffusion_eq}
\end{equation}
where 
$Q$ is the electron source strength, and 
$r$ is the distance to sources from the solar system.

%
%
We shall take into account the fact that 
the Galactic disk, where the sources are distributed, 
is surrounded by a halo in which the cosmic rays 
are confined for a long time 
before they escape into intergalactic space. 
We assume that 
cosmic-ray electrons are bounded by two parallel planes 
at $z={\pm}h$ with the median plane being occupied 
by the Galactic disk. 
The halo thickness, $h$, is estimated to be $2.8^{+1.2}_{-0.9}$kpc 
from $^{10}$Be observations from the HET aboard 
the Voyager 1 and 2 spacecraft \citep{lukasiak94}, 
which is also consistent with the observed Galactic radio emission structure 
\citep{beuermann85}. 
Thus we need to consider solutions 
with these boundary conditions for further analysis. 
Since the solar system is located at $z=15$ pc, 
quite near the median plane of the disk, 
and the sources are uniformly distributed in layer much thinner
than the confinement domain, 
we put $z = 0$. 
Assuming burst-like injection of electrons 
after the supernova explosion, 
the source term can be represented by 
\begin{equation}
 Q(E,r,t)=Q(E){\delta}(r){\delta}(t). 
\end{equation}
The density of electrons, $N_{\rm e}$, 
from a point source with injection spectrum 
\begin{equation}
 Q(E) = Q_0 E^{-\gamma} {\exp}(-E/E_{\rm c})
\end{equation}
at a distance $r$ and 
time $t$ after the release of electrons from the remnant,
is derived from the diffusion equation using the Fourier transform, 
assuming the diffusion coefficient has the form $D = D_{0}E^{\delta}$. 
Taking the boundary condition $N_{\rm e} = 0$ 
at the boundary of the Galactic halo $z = {\pm}h$, 
the general solution of the equation (\ref{eq:diffusion_eq}) 
in cylindrical coordinates is given by 
\begin{equation}
N_{\rm e}(E,r,t) = \sum_{n=0}^{\infty} \frac{Q_0}{4{\pi}D_{1}h}
 {\exp}(-D_{1}k_{n}^{2}-r^{2}/(4D_{1}))
 {\cos}(k_{n}z)(1-bEt)^{\gamma-2}E^{-\gamma}
 {\exp}(-E/(E_{\rm c}(1-bEt)))
 \label{eq:accurate_sol}
\end{equation}
where 
\[
 D_1 = {\int_{E}^{E/(1-bEt)}}(1/b)DE^{-2}dE 
     = \frac{D_0(1-(1-bEt)^{1-\delta})}{b(1-\delta)E^{1-\delta}}, 
\]
and 
\[
 k_{n} = \frac{\pi}{2h}(2n+1). 
\]

%
%
\subsection{Distant Components and Nearby Components}
\label{subsec:dist_comp}

Integrating $N_{\rm e}(E,r,t)$ of the solution (\ref{eq:accurate_sol}) 
with $r$ from 0 to $\infty$, 
we derive 
\begin{eqnarray}
 N_{\rm e}(E,t) &=& \int_{0}^{\infty}N_{\rm e}(E,r,t) 2{\pi}rdr \nonumber \\ 
       &=& \sum_{n=0}^{\infty}\frac{Q_0}{h}
 {\exp}(-D_{1}k_{n}^{2})(1-bEt)^{\gamma-2}E^{-\gamma}
 {\exp}(-E/(E_{\rm c}(1-bEt))). 
 \label{eq:cyl_net}
\end{eqnarray}
In this calculation, 
we approximated the integration on $r$ to range from 0 to $\infty$. 
Since electrons cannot propagate to larger distances than 
the halo thickness of $h$ in the radial direction, 
we may ignore the effects of the lateral distribution of the sources 
on the Galactic disk. 
In integrating $N_{\rm e}(E,t)$ of the formula (\ref{eq:cyl_net}) 
on $t$ from 0 to $1/bE$, 
we need the numerical integration. 
In the case of $E<<E_{\rm c}$, however, 
we can perform the integration with an analytical way, 
since the exponential cut-off term can be ignored. 
The solution is given by 
a Confluent Hyper-geometric Function for each term in the series. 
\begin{eqnarray}
 N_{\rm e}(E) &=& \int_{0}^{1/(bE)}f N_{\rm e}(E,t) dt \nonumber \\ 
             &=& \sum_{n=0}^{\infty}fQ_{0}
 \frac{_{1}F_{1}(1,(\gamma-\delta)/(1-\delta),-y)}
 {hb(\gamma-1)E^{\gamma+1}},  
\label{eq:cyl_sol}
\end{eqnarray}
where $_{1}F_{1}$ is a Confluent Hyper-geometric Function and 
\[
 y = D_{0}\frac{k_n^2}{(1-\delta)bE^{1-\delta}}, 
\]
and $f$ is the supernova explosion rate in the Galaxy 
per unit area and unit time. 
This analytic form is convenient to sum up many terms in the series. 
Higher energy electrons cannot reach the halo boundary 
due to their larger energy loss rate, 
and distribute around the Galactic disk plane. 
Therefore, 
we need to include the higher-order Fourier components 
to estimate the flux of higher energy electrons. 
In fact, we need to add up to 10 terms at 10GeV while up to 
$2\times10^{2}$ terms at 10TeV  
to get a precision of a few percent in the convergence 
of the series solution (\ref{eq:cyl_sol}). 
In order to simplify the calculation for the high energy region, 
we can use a three dimensional solution without boundary conditions 
instead of using the two dimensional solution (\ref{eq:cyl_sol}), 
which is given by 
(e.g. \citealt{berezinskii90} and other references therein) 
\begin{equation}
N_{\rm e}(E,r,t) = \frac{1}{(4{\pi}D_1)^{3/2}}e^{-r^2/(4D_1)}
 Q(\frac{E}{1-bEt})(1-bEt)^{-2}
 {\exp}(-E/(E_{\rm c}(1-bEt))), 
\label{eq:sphere_sol}
\end{equation}
where 
\[
 D_1 = {\int_{E}^{E/(1-bEt)}}(1/b)DE^{-2}dE. 
\]
The accuracy obtained by this approximation is discussed 
in Appendix \ref{sec:app_A}.

The flux of electrons from sources continuously and uniformly 
distributed in the Galactic disk is derived from the integration 
of electron density with respect to $r$ and $t$ in the following, 
\begin{equation}
J(E) = \frac{c}{4\pi}N_{\rm e}(E). 
\end{equation}
In the low energy region, below 1 TeV, 
many sources contribute to the observed electron flux.
As the energy of the observed electrons increases, 
the number of electron sources decreases and 
only nearby sources can contribute to the electron flux. 
As described in section \ref{subsec:diffusion},  
young cosmic-ray electron sources which are nearby 
create discrete effects 
in the TeV region, such as 
structure in the electron spectrum and 
the anisotropy favoring the source direction. 
Thus we can define the contribution from distant or relatively old sources 
by subtracting the flux of nearby young sources, 
eliminating the effects of fluctuations due to the small number of nearby sources, 
as follows: 
\begin{equation}
 J_{\rm d}(E) = J(E) - J_{\rm n}(E), 
\end{equation}
where $J_{\rm d}$ and $J_{\rm n}$ show 
the flux of electrons from distant (in space-time) sources 
and nearby sources, respectively. 
Here, we calculated the contribution of nearby sources 
distributed uniformly in space and time as 
\begin{equation}
J_{\rm n}(E) = \frac{c}{4\pi} 
 \int_{0}^{T_0}dt \int_{0}^{R_0}dr f N_{\rm e}(E,r,t) 2{\pi}r. 
 \label{eq:Jlocal}
\end{equation}
Appropriate values of $R_0$ and $T_0$ are discussed 
in section \ref{subsec:spec2data}.

\section{SNRs as the Cosmic-Ray Electron Sources}
\label{sec:snrs}

%
%
As described in section \ref{sec:intro}, 
recent X-ray and TeV gamma-ray observations indicate that 
high energy electrons are accelerated in SNRs. 
Here, 
we assumed that SNRs are cosmic-ray electron sources, 
and that electrons are liberated burst-likely from SNRs 
at the time of ${\tau}=0$, $5{\times}10^3$~yr, 
$1{\times}10^4$~yr, $5{\times}10^4$~yr, and
$1{\times}10^5$~yr after the explosion. 
As described in section \ref{sec:result}, 
the case of the continuous release is well approximated by 
the burst-like release with a mean value of the liberation time.

%
%
From the observations of SN1006 
\citep{reynolds96, koyama95, tanimori98}, 
the output energy in electrons above 1 GeV is estimated 
to be $\sim1\times10^{48}$ erg. 
We can also estimate the net output energy for electrons from supernovae 
using the energy density of cosmic rays and the supernova rate 
of $1/30{\rm yr}$ 
in the Galaxy (\citealt{berezinskii90, gaisser00} 
and other references therein). 
The output energy of electrons above 1 GeV 
estimated in this way is about $10^{48}$ erg/SN, 
and is consistent with the estimate 
from SN1006. 
Therefore, we took the output energy from a supernova in electrons 
above 1 GeV is $W=1{\times}10^{48}$ erg.

%
%
To analyze contributions due to discrete cosmic-ray electron sources, 
we list in Table \ref{tab:nearbysnrs} all known SNRs
which are located within  1~kpc of the solar system 
and with age less than $4\times10^5$~yr. 
For each source, the maximum energy $E_{\rm max}$ of electrons 
reaching the solar system is given by $E_{\rm max}=1/(bT)$, 
with the energy loss coefficient $b$ shown in figure \ref{fig:enelossb} 
and the age $T$ of each SNR, 
which are determined independent of the diffusion coefficient.

\begin{center}
\vspace*{0.5cm}
\framebox[3cm]{Table \ref{tab:nearbysnrs}}
\vspace*{0.5cm}
\end{center}

Distance to SNRs is an important parameter in evaluating 
the high energy electron flux, 
and we comment on this parameter of some SNRs 
listed in Table \ref{tab:nearbysnrs}.

So far, the canonical distance to the Vela SNR has been taken
to be 500pc, a value which was derived from 
the analysis of its angular diameter in comparison with the
Cygnus Loop and IC443 \citep{milne68}, 
and pulsar dispersion determination \citep{taylor93}. 
However, recent parallax measurements clearly indicate 
that the distance of 500pc is too large. 
\citet{cha99} obtained high resolution Ca-II absorption line 
toward 68 OB stars in the direction of the Vela SNR. 
The distances to these stars were determined by 
trigonometric parallax measurements with the Hipparcos satellite and 
spectroscopic parallaxes based upon photometric colors and spectral types. 
The distance to the Vela SNR is constrained to be $250\pm30$pc due to 
the presence of the Doppler spread Ca-II absorption line  
attributable to the remnant along some lines of sight. 
\citet{caraveo01} also applied high-resolution astrometry to 
the Vela pulsar (PSR B0833-45) $V\sim23.6$ optical counterpart. 
Using Hubble Space Telescope observations, 
they obtained the first optical measurement of the annual parallax 
of the Vela pulsar, yielding a distance of $294^{+76}_{-50}$ pc. 
Therefore, 
we calculate the electron flux adopting a distance of 
300~pc to the Vela SNR.

Previously, a distance of 770 pc to the Cygnus Loop 
was often quoted \citep{minkowski58}. 
Recently \citet{blair99} suggested that the distance is 
$440^{+130}_{-100}$ pc, 
based on Hubble Space Telescope observations of a filament 
in the remnant. 
Thus we took the distance to be 440 pc, 
considerably smaller than the previously quoted distance of 770 pc.

For the recently discovered SNR RX J0852.0-4622, 
the estimated age of $\sim680$yr and 
distance of $\sim200$pc was based on 
thermal X-ray emission and $^{44}$Ti gamma-ray detection 
\citep{aschenbach98, iyudin98}.  
However, 
ASCA observations reveal that the X-ray emission is non-thermal, 
and re-analysis of the COMPTEL data indicates that 
the $^{44}$Ti gamma-ray detection is only significant 
at the $2-4{\sigma}$ level (\citealt{slane01} and references therein). 
Thus it is suggested that RX J0852.0-4622 is at a larger distance, 
 $1-2$kpc \citep{slane01}, 
so we do not include it in Table \ref{tab:nearbysnrs}.

\section{Results}
\label{sec:result}

\subsection{Electrons from the Known Nearby SNRs}

\begin{figure}
\epsscale{0.70}
\plotone{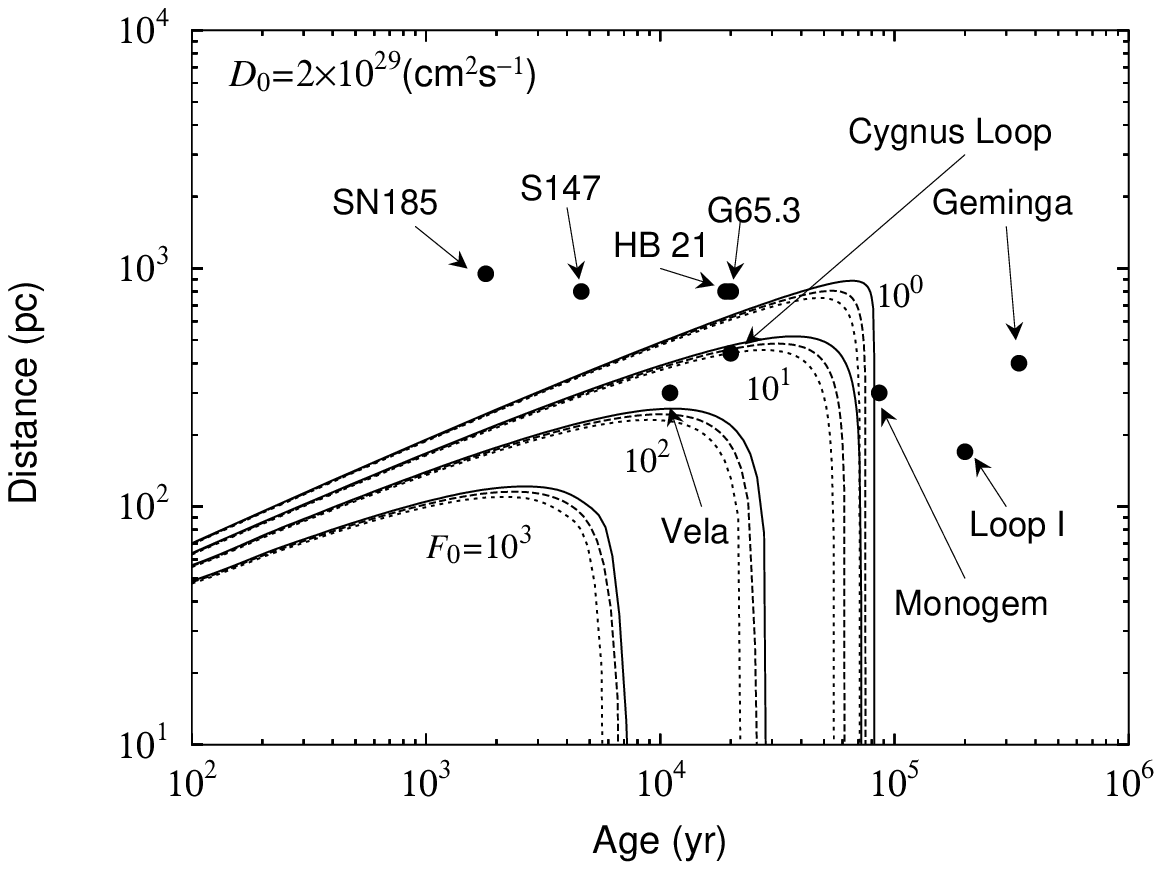}

\epsscale{0.70}
\plotone{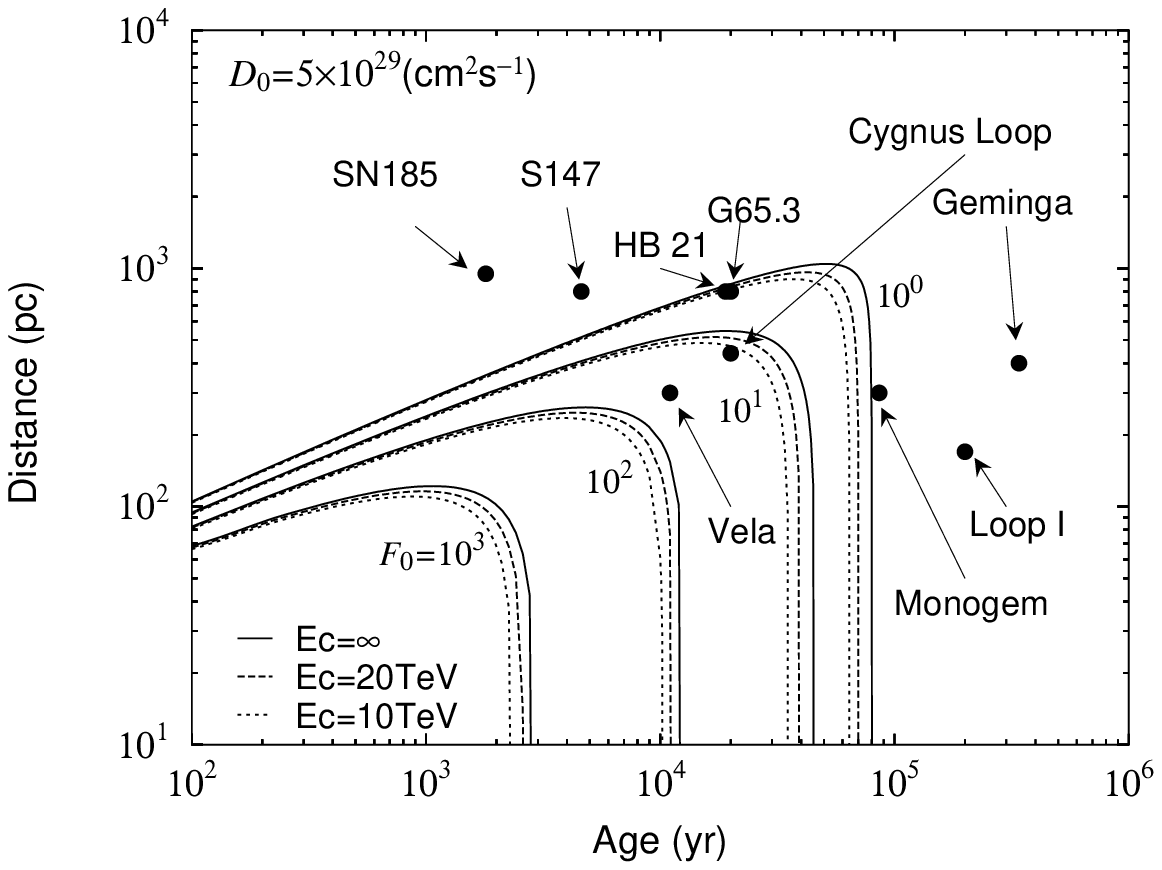}
\caption{\label{fig:age_distance}
Contours of the electron flux at 3 TeV between distances and ages 
with the values of each SNR, 
in the case of the prompt release of electrons after the explosion. 
Lines show equal flux contour for 
$F_0 = (E/{\rm GeV})^3J$(GeV$^2$m$^{-2}$s$^{-1}$sr$^{-1}$), 
where $J$ is the flux of electrons at 3 TeV. 
Contour levels are $10^3$, $10^2$, $10^1$, and 
$10^0$ (GeV$^2$m$^{-2}$s$^{-1}$sr$^{-1}$) 
with the output energy of electrons over 1GeV of 
$W=1\times10^{48}$ erg/SN. 
Here, the injection spectrum is a power-law with an exponential cut-off of 
$E_{\rm c}$=10~TeV (dotted line), 20~TeV (dashed line), 
and $\infty$ (solid line). 
Upper and Lower panels show the flux contours 
for $D=D_{0}(E/{\rm TeV})^{0.3}$ with 
$D_0 = 2\times10^{29}$cm$^2$s$^{-1}$ 
and $D_0 = 5\times10^{29}$cm$^2$s$^{-1}$, respectively.  
}
\end{figure}

%
%
Figure \ref{fig:age_distance} shows contours of 
the expected electron flux at 3 TeV 
(with flux values scaled by $E^3$ ) 
as a function of age and distance of the SNR. 
As described in the preceding section, 
we assumed that the output energy of electrons over 1 GeV 
is $W=1\times10^{48}$ erg/SN, and 
the injection spectrum is a power-law function with 
an exponential cut-off of the form of ${\exp}(-E/E_{\rm c})$. 
The diffusion coefficient has the form 
$D=D_{0}(E/{\rm TeV})^{\delta}$ cm$^2$s$^{-1}$, 
where $\delta$ is 0.3 in the TeV region and 
the spectral index $\gamma$ is chosen to maintain
$\gamma+\delta=2.7$. 
As shown in Figure \ref{fig:age_distance}, 
the electron flux is strongly dependent on source age and distance. 
In the case of the prompt release of electrons after the explosion, 
the flux from the Vela SNR is the largest among the known SNRs 
listed in Table \ref{tab:nearbysnrs}.  
The flux value is quite sensitive to the 
change of distance to Vela from 500pc to 300pc, 
since the solution for electron density yields 
a Gaussian distribution function of $r$  
as shown in formula (\ref{eq:accurate_sol}) or (\ref{eq:sphere_sol}). 
The flux of electrons at a distance of 300pc 
is two orders of magnitude larger than at 500pc.

\begin{center}
\vspace*{0.5cm}
\framebox[3cm]{Figure \ref{fig:age_distance}}
\vspace*{0.5cm}
\end{center}

\subsection{Calculated Electron Energy Spectrum Compared with the
  Observed Spectrum}
\label{subsec:spec2data}

We separately calculated the contributions to the electron energy spectrum 
from nearby and distant sources.
We selected SNRs in the neighborhood of the solar system located within
a distance of 1kpc and an electron release time within $10^5$yr 
in the past. 
The boundary condition in this domain almost corresponds to 
the diffusion path length and life time of electrons around 1~TeV. 
We calculated the electron flux from nearby sources using the 
distance and ages of selected SNRs 
in Table \ref{tab:nearbysnrs} with formula (\ref{eq:sphere_sol}). 
Justification of taking $R_{0}=1$kpc and $T_{0}=10^{5}$yr is 
discussed in Appendix B. 
The SNRs in the nearby region are thus
SN185, S147, HB 21, G65.3+5.7, Cygnus Loop, Vela, and Monogem. 
In this calculation, we assumed that 
supernovae occur uniformly on the Galactic disk  
at the rate of 1/30 yr, and took the halo thickness to be $h=3$kpc.

\begin{figure}
\epsscale{0.80}
\plotone{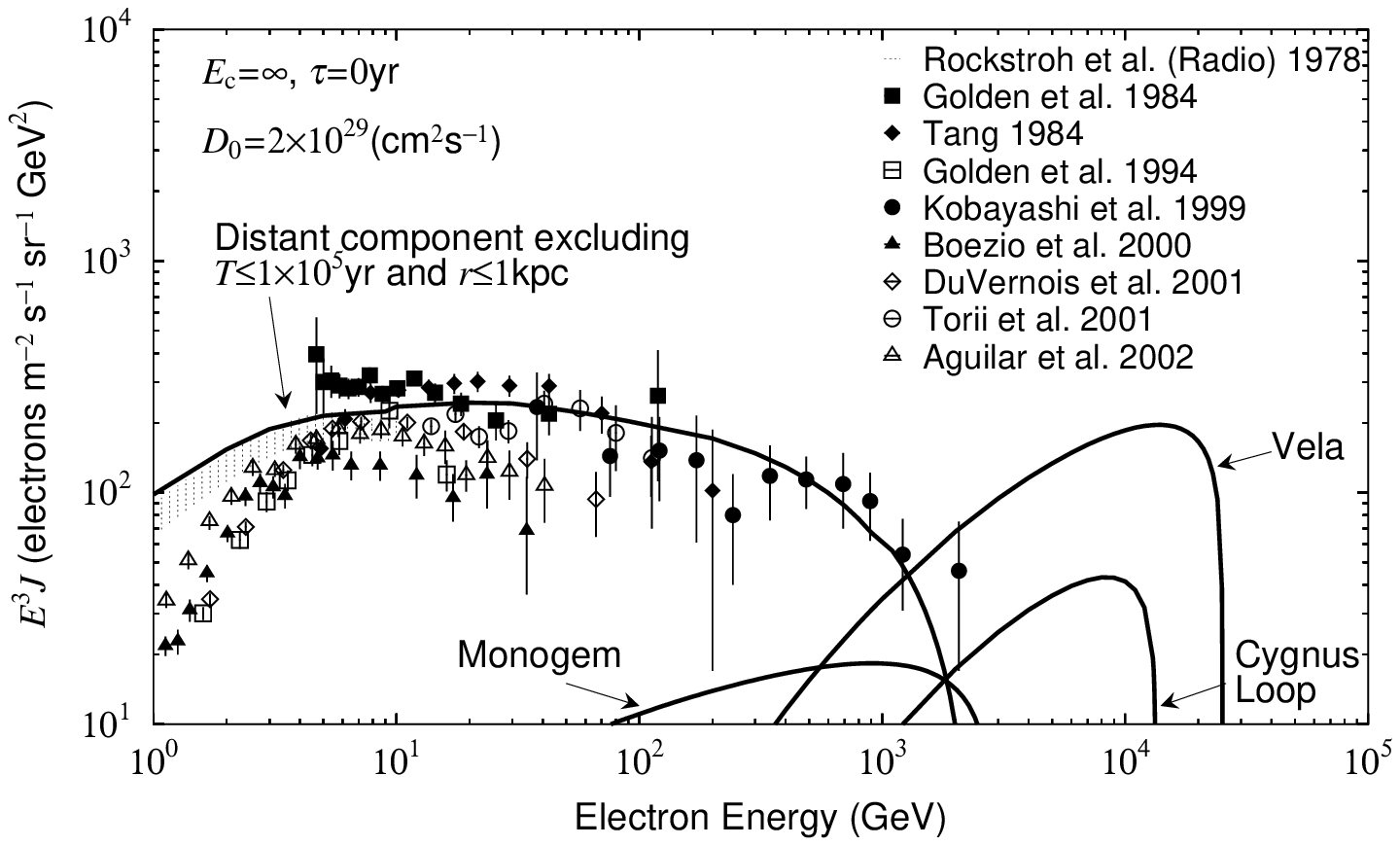}

\vspace{5mm}

\epsscale{0.80}
\plotone{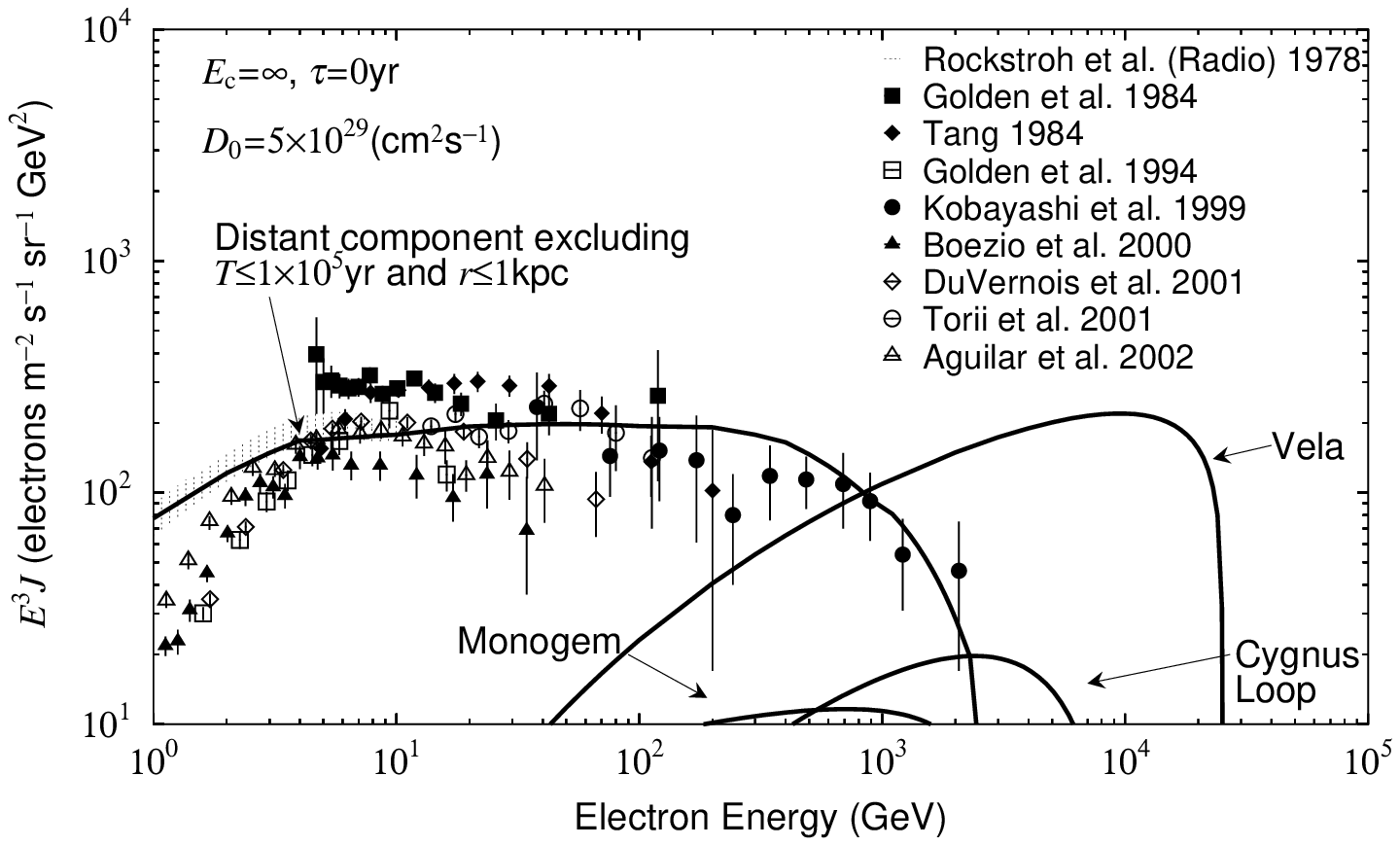}

\caption{\label{fig:elespec}
Calculated energy spectra of electrons without a cut-off 
of the injection spectrum 
for the prompt release after the explosion ($\tau=0$), 
compared with presently available data. 
Here we took the diffusion coefficient of  
$D=D_{0}(E/{\rm TeV})^{0.3}$ with 
$D_{0} = 2\times10^{29}$cm$^2$/s and $D_{0} = 5\times10^{29}$cm$^2$/s 
in TeV region, 
and $D=2{\times}10^{28}(E/{\rm 5GeV})^{0.6}$ in GeV region 
as given by the formula (\ref{eq:lowD}). 
See text for details. 
}
\end{figure}

Figure \ref{fig:elespec} shows the calculated energy spectra of
electrons without a cut-off of the injection spectrum in the case of 
the prompt release after the explosion ($\tau=0$), 
compared to the observed data 
\citep{golden84, tang84, golden94, kobayashi99, boezio00, alcaraz00,  
duvernois01, torii01}. 
Here, we illustrate the cases of the diffusion coefficient of 
$D = D_{0}(E/{\rm TeV})^{0.3}$ with  
$D_0 = 2\times10^{29}$ cm$^2$s$^{-1}$ and 
$D_0 = 5\times10^{29}$ cm$^2$s$^{-1}$ in TeV region, 
with $D = 2\times10^{28}(E/{\rm 5GeV})^{0.6}$ cm$^2$s$^{-1}$ 
upto 50~GeV or 1~TeV as given by the formula ({\ref{eq:lowD}}) 
in section \ref{subsec:diffusion}. 
In this figure, we also plotted the flux of low energy interstellar 
electrons estimated from the Galactic radio data 
\citep{rockstroh78}. 
Cosmic-ray electrons have been observed by a variety of instruments, 
but only emulsion chamber data provide observations of 
electrons above 1 TeV.
Since the flux of TeV electrons is low, detectors must have
large acceptance $S{\Omega}$ and high proton rejection power ($\sim10^5$). 
The emulsion chamber satisfies these requirements 
\citep{nishimura80, kobayashi99}. 

Our propagation model, 
which consists of separately calculated distant and nearby components, 
is consistent with the observed data for the local primary electron spectrum 
in the energy range from 10GeV to 2TeV. 
Our model also predicts that nearby SNRs insert unique, 
identifiable structures in the electron spectrum from 1TeV to 10TeV. 
As shown in Figure \ref{fig:elespec}, 
the absolute flux and spectral shape change with the diffusion coefficient. 
On the other hand, the maximum energy of each SNR is the same, 
independent of the diffusion coefficient value, 
because it is determined by the age of the SNR.

\begin{center}
\vspace*{0.5cm}
\framebox[3cm]{Figure \ref{fig:elespec}}
\vspace*{0.5cm}
\end{center}

\begin{figure}
\epsscale{0.80}
\plotone{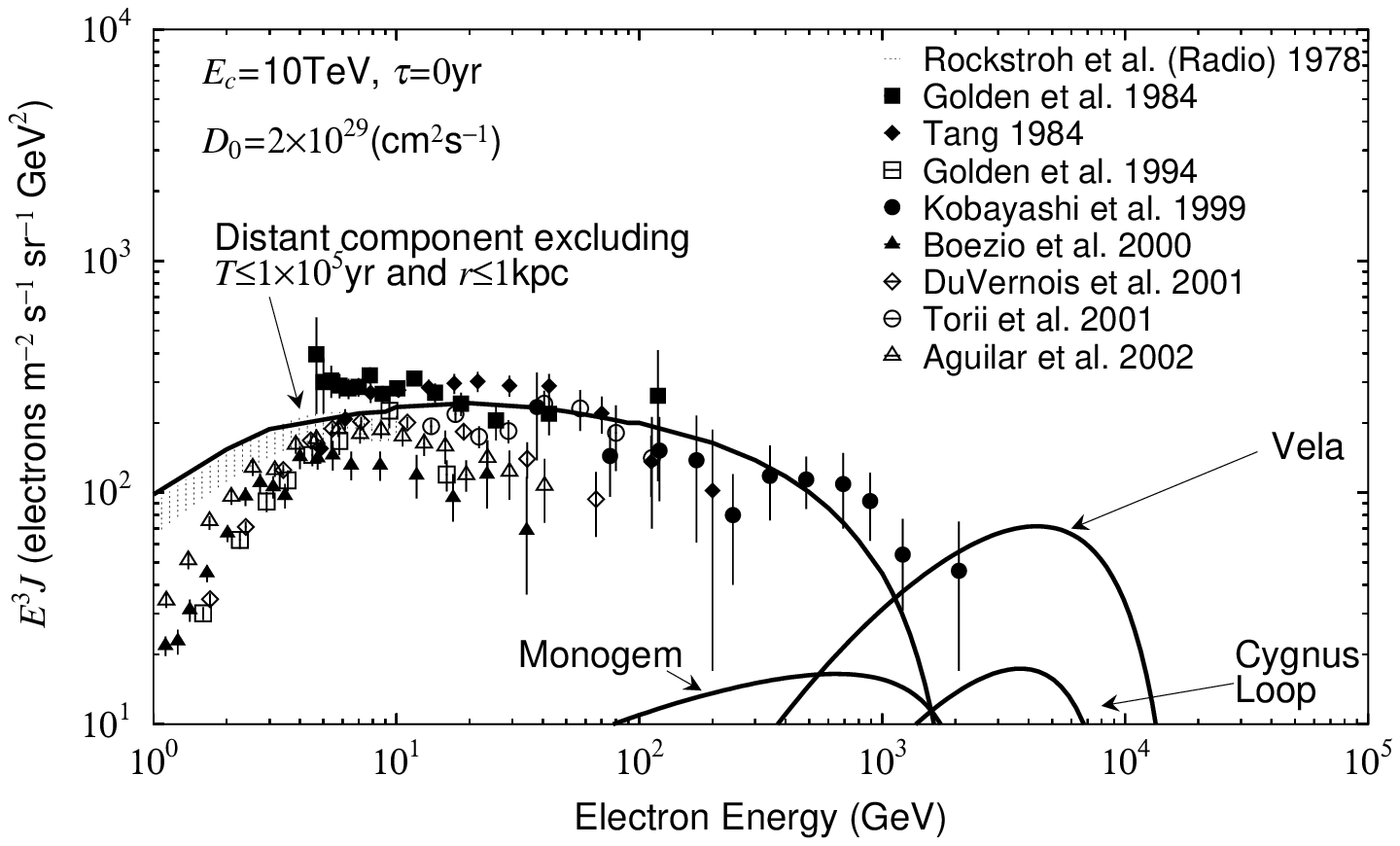}

\vspace{5mm}

\epsscale{0.80}
\plotone{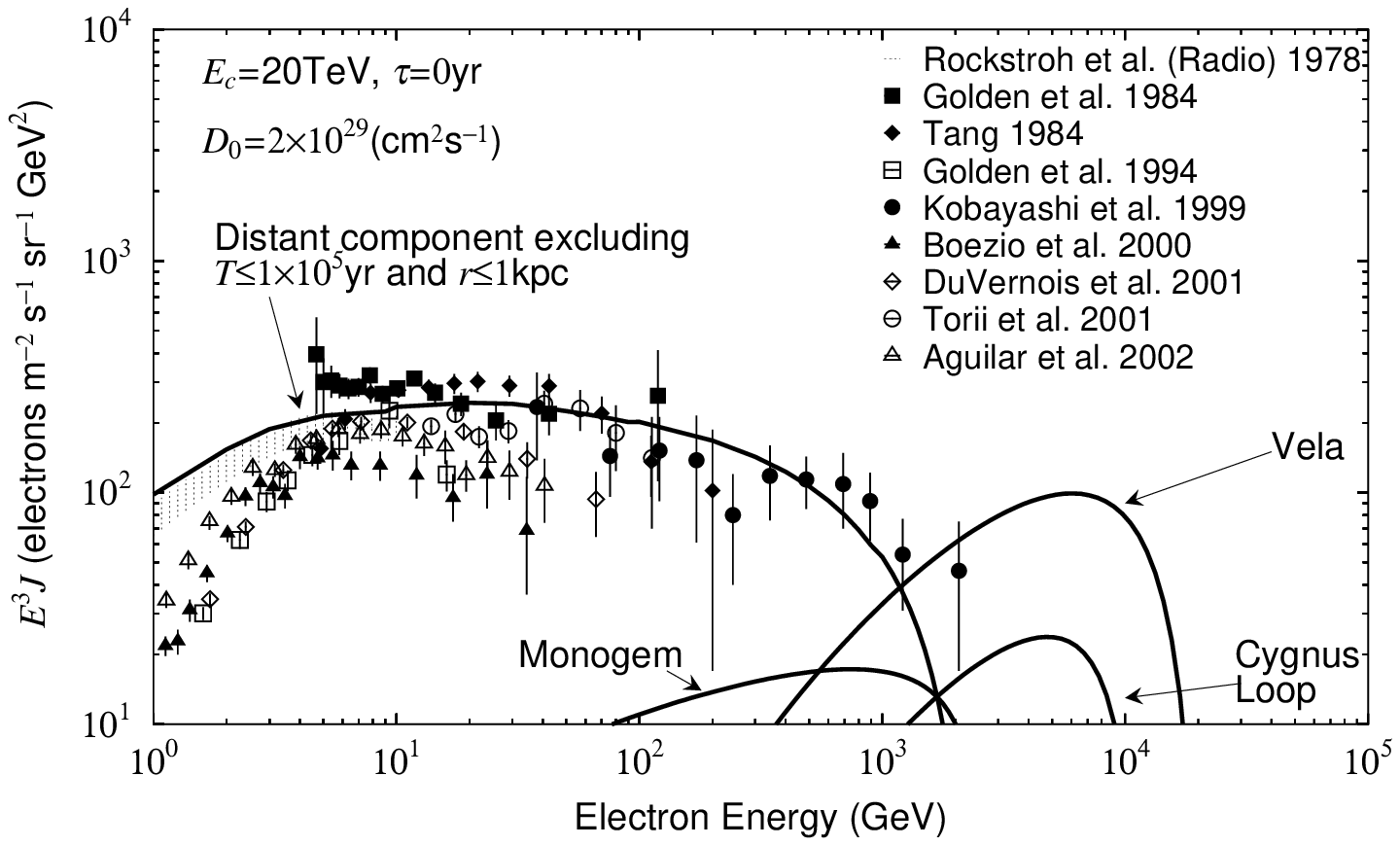}

\caption{\label{fig:elespec_cutoff}
Calculated energy spectra with a cut-off of $E_{\rm c}$=10~TeV 
and 20~TeV for the prompt release after the explosion ($\tau=0$). 
Here we took the diffusion coefficient of 
$D_{0} = 2\times10^{29}$cm$^2$/s. 
}
\end{figure}

\begin{figure}
\epsscale{0.50}
\plotone{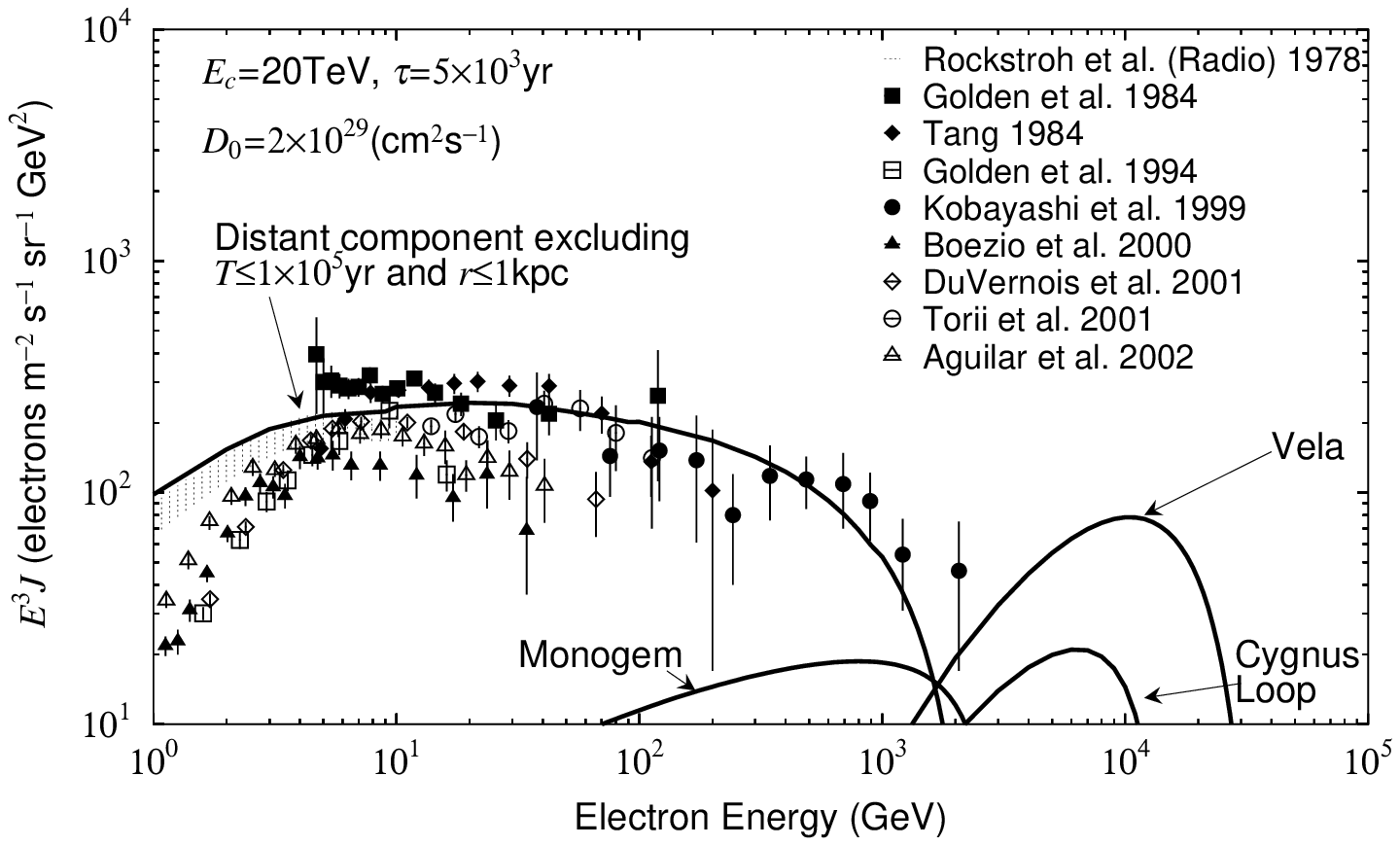}
\epsscale{0.50}
\plotone{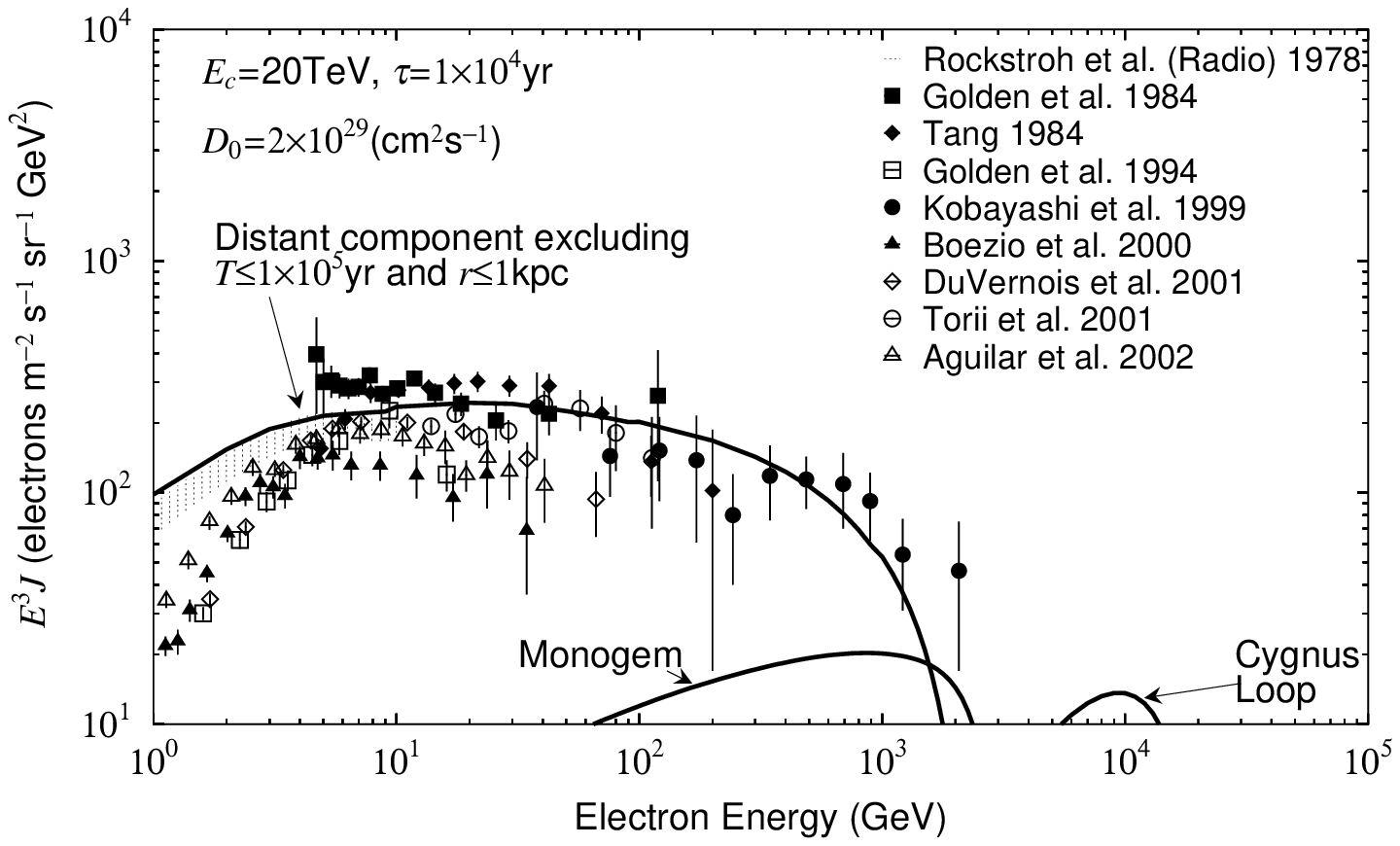}
\epsscale{0.50}
\plotone{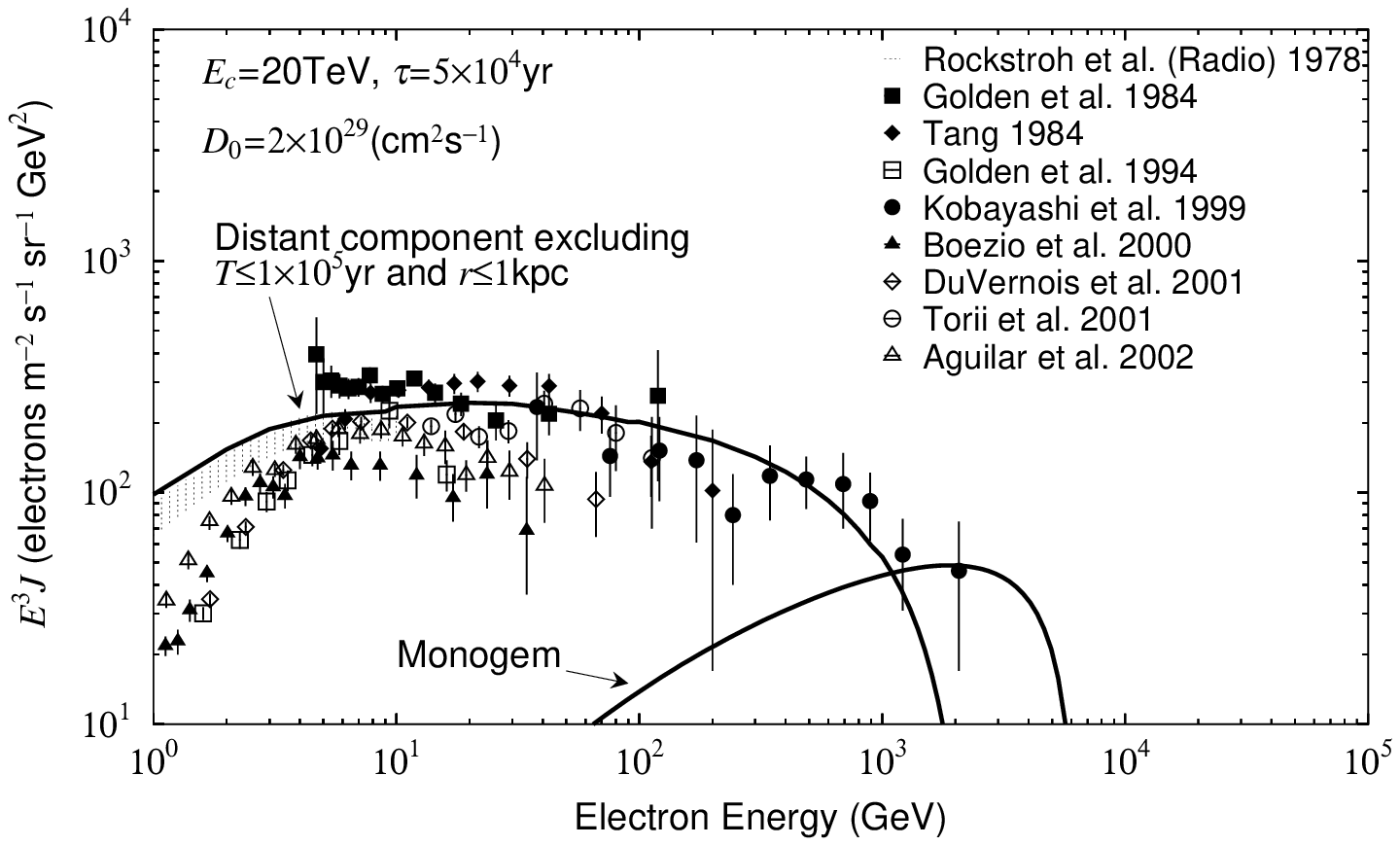}
\epsscale{0.50}
\plotone{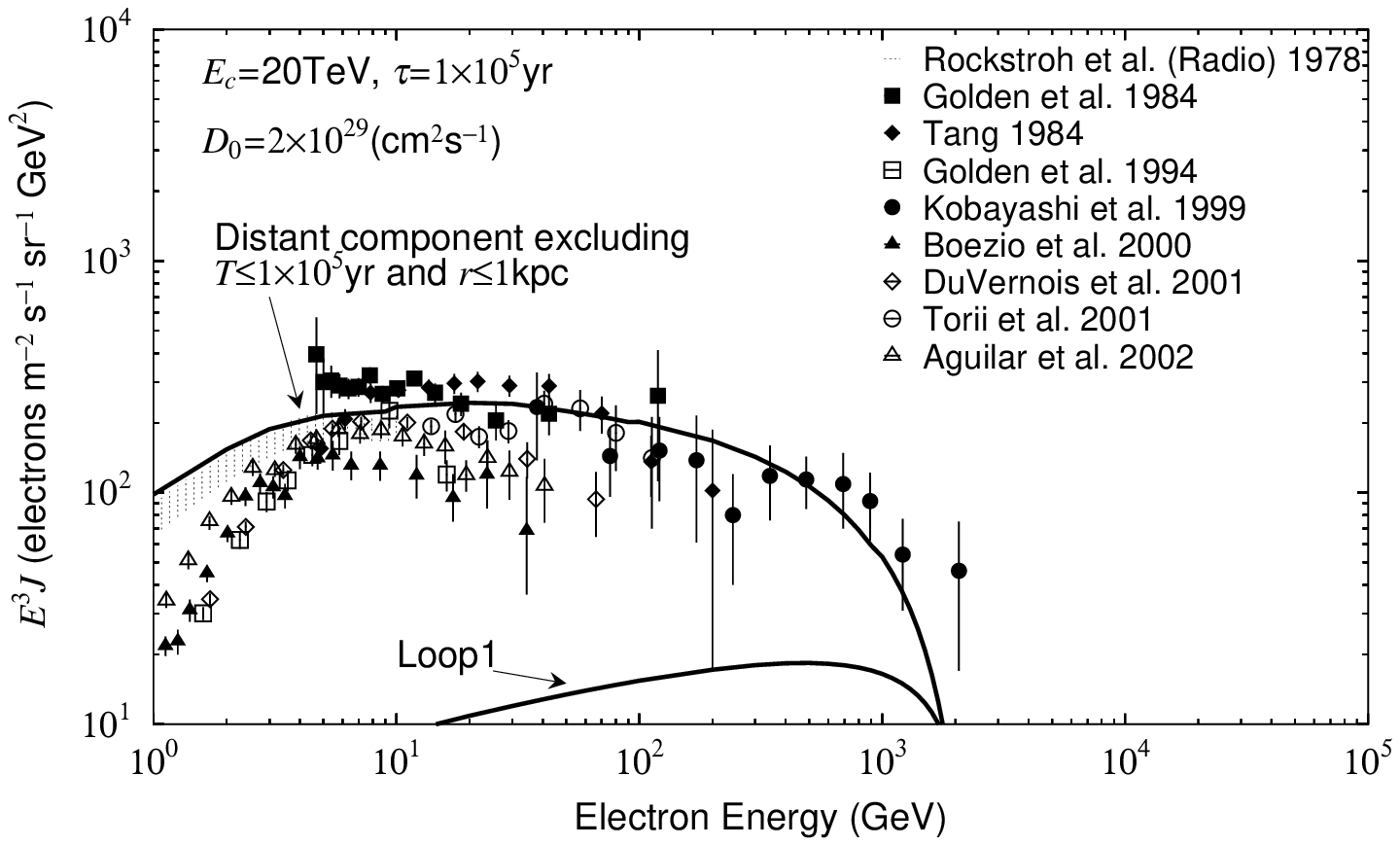}
\caption{\label{fig:elespec_tau}
Calculated energy spectra with 
$D_{0} = 2\times10^{29}$cm$^2$/s and a cut-off of $E_{\rm c}$=20~TeV 
for the burst-like release at $\tau=5{\times10^3}$~yr, 
$1{\times10^4}$~yr, $5{\times10^4}$~yr, and $1{\times10^5}$~yr, respectively. 
}
\end{figure}

We also calculated how the source spectral shape and 
release time of electrons 
from the remnant affect the cosmic-ray electron spectrum. 
We illustrate some examples of these calculations by taking 
$D_{0} = 2\times10^{29}$ cm$^2$s$^{-1}$. 
Figure \ref{fig:elespec_cutoff} shows the calculated energy spectra 
with a cut-off of $E_{\rm c}=$10~TeV, 20~TeV, and $\infty$ 
in the injected electron spectrum, 
assuming the prompt release after the explosion. 
We can find that these spectra are similar with each other, 
independent of the cut-off energies. 
Figure \ref{fig:elespec_tau} shows the calculated energy spectra 
with a cut-off of $E_{\rm c}=20$TeV, 
in which electrons are released burst-likely after the explosion 
in the release time of $\tau=5\times10^3$yr, $1\times10^4$yr, 
$5\times10^4$yr, and $1\times10^5$yr, respectively. 
As shown in this figure, 
it is to be noted that the spectrum for ${\tau}=5{\times}10^{3}$yr 
is almost the same with that of the prompt release after the explosion. 
However, the delay of the release time from SNRs 
have a large impact on the flux in the TeV region for 
$\tau{\geq}1{\times}10^{4}$yr.

\begin{center}
\vspace*{0.5cm}
\framebox[3cm]{Figure \ref{fig:elespec_cutoff}}
\vspace*{0.5cm}
\end{center}

\begin{center}
\vspace*{0.5cm}
\framebox[3cm]{Figure \ref{fig:elespec_tau}}
\vspace*{0.5cm}
\end{center}

\begin{figure}
\epsscale{0.50}
\plotone{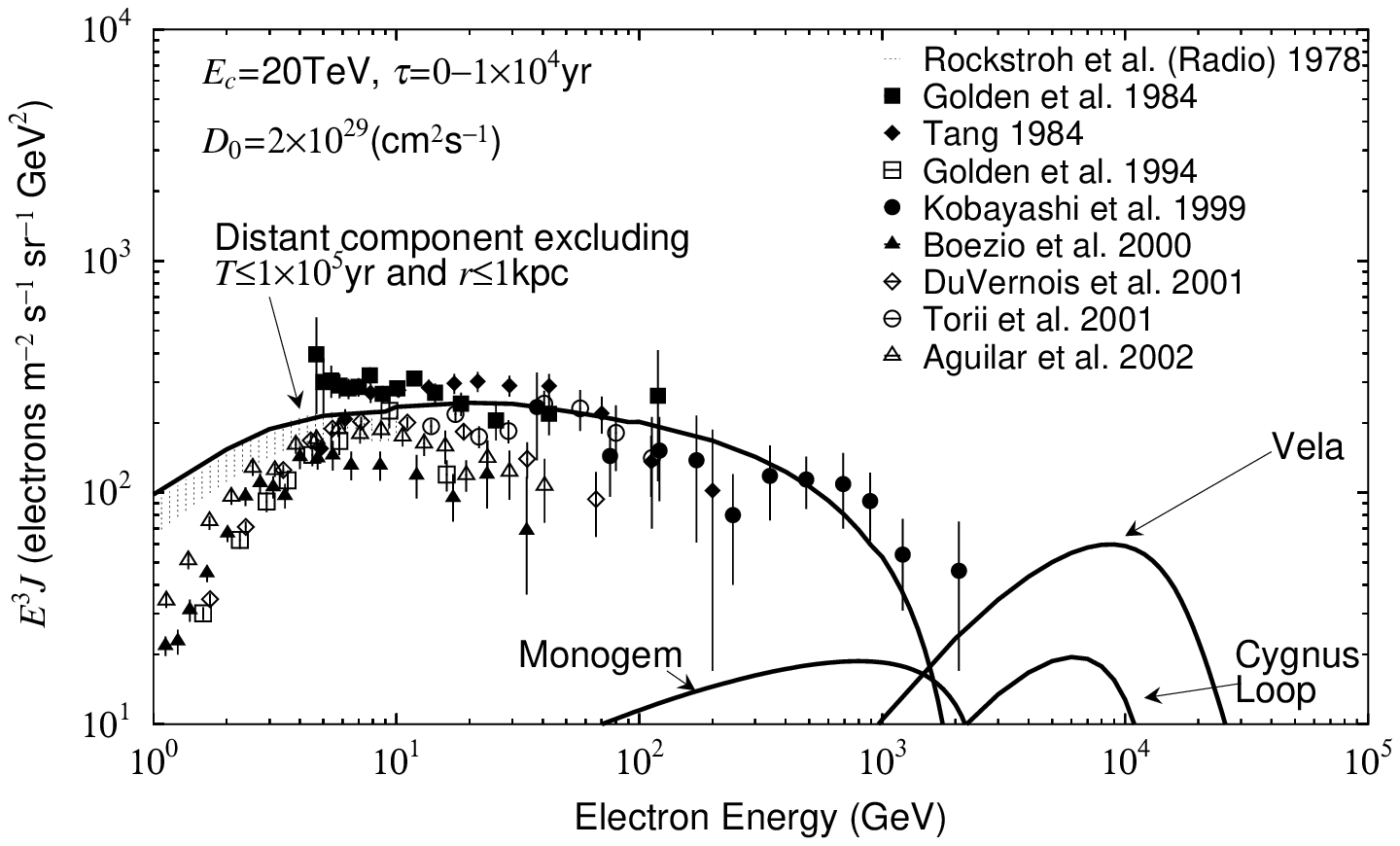}
\epsscale{0.50}
\plotone{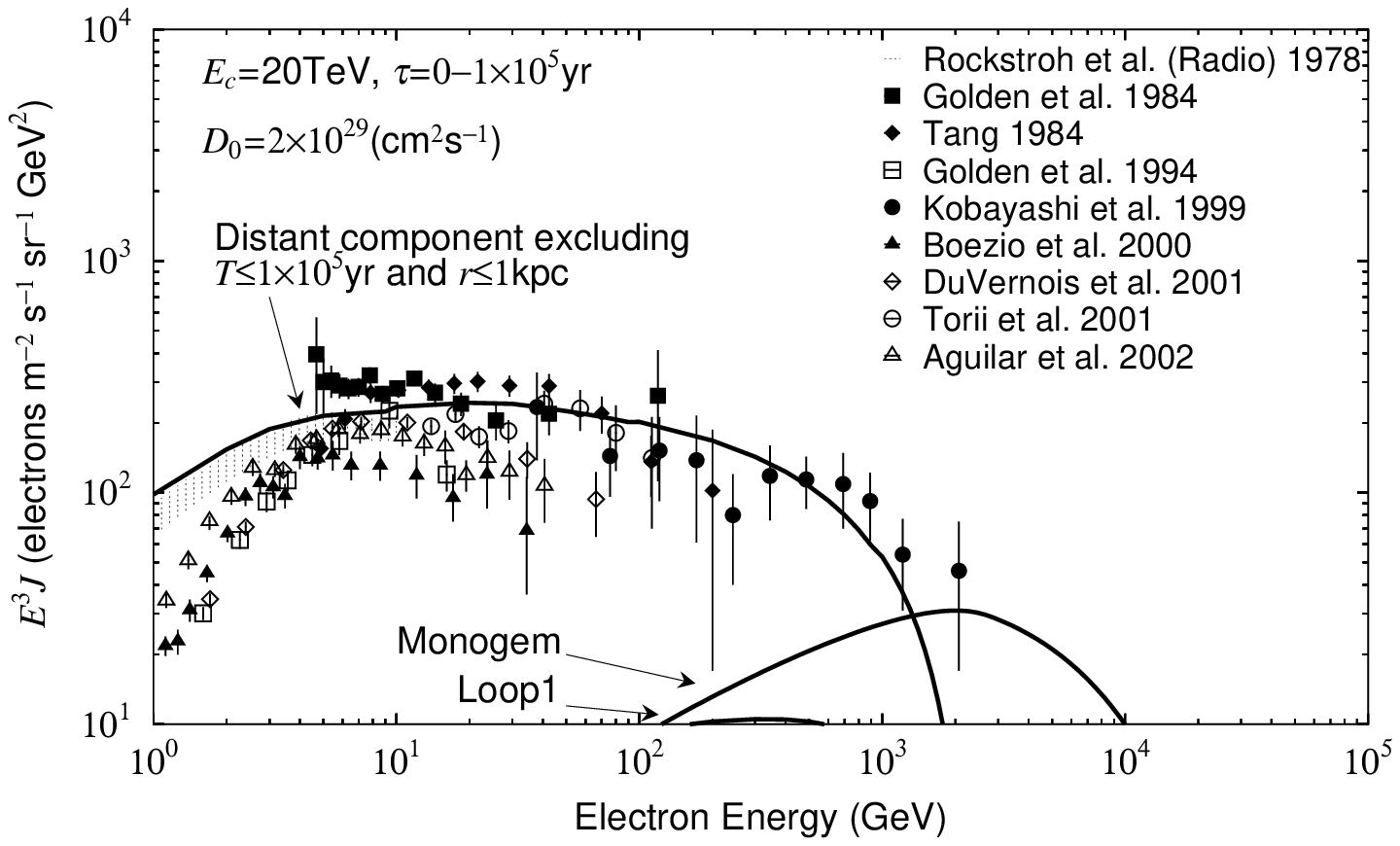}
\caption{\label{fig:elespec_cont}
Calculated energy spectra with 
$D_{0} = 2\times10^{29}$cm$^2$/s and a cut-off of $E_{\rm c}$=20~TeV 
for the continuous release time of ${\tau}=0{\sim}1{\times10^4}$~yr 
and ${\tau}=0{\sim}1{\times10^5}$~yr, respectively. 
}
\end{figure}

We also checked the case of the continuous release. 
Figure \ref{fig:elespec_cont} shows 
that these spectra are well represented by that of 
the burst-like release with a mean value of the continuous release time.

\begin{center}
\vspace*{0.5cm}
\framebox[3cm]{Figure \ref{fig:elespec_cont}}
\vspace*{0.5cm}
\end{center}

\section{Summary and Discussion}

%
%
We calculated the energy spectrum of cosmic-ray electrons, 
separating the contributions of 
sources nearby in space and time (those within distances of 1kpc and 
times of $1\times10^5$yr) and Galactic sources 
located outside this domain.

%
%
There are only 9 SNRs 
within a distance of 1kpc and an age of $4\times10^{5}$yr,
as shown in Table \ref{tab:nearbysnrs}, 
which is much smaller than the expected number of $\sim60$ 
assuming a supernova rate of 1/30yr and 
a Galactic disk radius of 15kpc. 
This is due to selection biases in radio observations, 
since surface brightnesses in radio observations decrease with age.
Thus, also coupled with the reason of the arguments in Appendix B, 
we define a domain of $R\leq$1kpc and $T\leq 1\times10^{5}$yr 
to evaluate the electron flux from nearby sources. 
There are 7 known SNRs 
within this region listed
in Table \ref{tab:nearbysnrs}:
Vela, Cygnus Loop, Monogem, G65.3+5.7, HB 21, S147, and SN185. 
On the other hand, 
the expected number of SNRs in this smaller domain is $\sim15$. 
The difference could be partly due to statistical fluctuations in the small 
number of sources, 
and it could also be due to undetected SNRs in this domain. 
In fact, the surface brightness estimated from 
an age of $1\times10^{5}$yr, assuming adiabatic phase, 
is still fainter than the typical detection limit of surface brightness 
in studies of the distribution of Galactic SNRs 
\citep{kodaira74, leahy89}.
Although this indicates that the electron flux calculated 
from known nearby SNRs may give a lower limit of the flux,  
the contribution of undetected SNRs is expected to be relatively small 
in the case of the prompt release of electrons after the explosion. 
Since these undetected SNRs should have lower surface brightness and 
be older than detected SNRs, 
they could scarcely contribute to the electron intensity 
in the TeV region as indicated in figure \ref{fig:age_distance}. 
Some quantitative arguments on this point are discussed in Appendix B.

Our model predicts that nearby SNRs such as 
the Vela, Cygnus Loop, or Monogem, present unique, 
identifiable structures in the electron spectrum from 1TeV to 10TeV. 
As the diffusion coefficient increases, 
electrons propagate larger distances with longer mean path length, 
resulting the electron density becomes smaller. 
This causes a flatter spectral shape and smaller peak flux, 
as shown in figures \ref{fig:elespec}. 
The release time $\tau$ after the explosion determines 
which SNRs contribute for cosmic-ray electrons in the TeV region. 
For $\tau=0$ and $\tau=5\times10^3$yr, 
the energy spectra are similar with each other, and 
the Vela SNR is the most dominant source in the TeV region. 
For $\tau=1\times10^4$yr and $\tau=5\times10^4$yr, 
the Cygnus Loop and the Monogem SNR 
are dominant, and 
for $\tau=1\times10^5$yr 
there are no dominant known sources in the TeV region. 
We can see that some combinations of parameters 
(e.g. $D = 5\times10^{29}(E/{\rm TeV})^{0.3}$ cm$^2$s$^{-1}$, 
${\tau}=0$, and $E_{\rm c}={\infty}$) 
are already ruled out even by the limited existing data.

%
%
Besides such primary electrons accelerated in SNRs, 
the observed cosmic-ray electron flux includes 
secondary electrons, produced mainly by interactions of cosmic-ray 
protons and nuclei with interstellar gas. 
The secondaries are mostly decay products of 
charged pions produced in interactions, {\it i.e.} 
${\pi}^{\pm} \rightarrow {\mu}^{\pm} \rightarrow e^{\pm}$. 
As positrons are produced by this process, 
we can estimate the intensity of secondaries relative to primaries 
to be $\sim10$\% at 10 GeV 
from the observed ratio of $e^{+}$ to $e^{-}+e^{+}$  
$\sim5$\% at 10 GeV \citep{barwick97, boezio00}. 
The spectral index $\gamma=2.7$ for secondaries is the same as
the index of the parent cosmic-ray protons and nuclei. 
This index is larger than that of primary electrons 
at the source, which is $\sim$2.4 in the TeV region. 
Since the intensity relative to primary electrons 
decreases with increasing energy as  
$(E/10{\rm GeV})^{{\sim}2.4-2.7} = (E/10{\rm GeV})^{{\sim}-0.3}$,  
the relative intensity of secondaries becomes  $\sim$2.5\% at 1 TeV. 
This suggests that 
the intensity of the secondary electrons is negligible  
compared to primary electrons in the energy range of interest here.

%
%
Next, we estimate the degree of anisotropies from the specific sources 
in TeV region. 
Let $N_i(E,\mathbf{r}_i, t_i)$ be the contribution 
to the local density of cosmic-ray electrons 
at energy $E$ from a source located at distance $\mathbf{r}_i$ 
and of age $t_i$. 
The anisotropy parameter $\Delta_i$ due to the density gradient 
of electrons is given by 
\begin{equation}
 \Delta_i \equiv 
  \frac{I_{{\rm max}}-I_{{\rm min}}}{I_{{\rm max}}+I_{{\rm min}}} 
  = \frac{3D}{c}\frac{{\nabla}N_i}{N_i}
  = \frac{3r_i}{2ct_i} 
\end{equation}
for an individual source, 
where $I_{\rm max}$ and $I_{\rm min}$ are the 
maximum and minimum electron intensity in all directions 
\citep{shen71, ptuskin95}. 
For example, 
the anisotropy of electrons for Vela with $\tau=0$ is estimated to be 13\%,  
where $r_i = 300$pc and $t_i = 1.1\times10^4$yr. 
The anisotropy becomes larger if we include the effect of the delay 
of the release time from SNRs.

%
%
In this paper, it is demonstrated that 
measurements of the energy spectrum of electrons in the TeV region 
are crucial to detect the unique effects of nearby sources as described above. 
If we observe pronounced features in the shape of the spectrum, 
together with anisotropy towards the nearby SNRs, 
we would confirm the nearby SNRs as the main contributors 
to electrons in the TeV region. 
We can also perform a more detailed analysis for 
the release time of electrons, cut-off energy, output energy, 
diffusion coefficient, and so on. 

High energy cosmic-ray electrons are the most powerful probe 
to help us identify the origin of cosmic rays. 
However, at present, only emulsion chamber data are available 
in the energy region $1\sim 2$TeV. 
To make significant additional observations on electrons in this region, 
not only balloon flight experiments with long exposures 
but also new experimental programs such as 
the CALET (CALorimetric Electron Telescope)
on the International Space Station \citep{torii02} 
are to be promising to observe the electron spectrum 
and anisotropy in TeV region 
with high statistical accuracy. 
Such direct observations will reveal the origin of 
cosmic-ray electrons in this energy region, and also 
bring us important information on the sources, 
acceleration, and propagation of cosmic-ray electrons.

\acknowledgments

We are grateful to Prof. R.J.Wilkes 
for his careful reading and useful comments of the manuscript.

\appendix

\section{Relation between the three dimensional solution of the diffusion equation
without boundaries, and the two dimensional solution with boundaries}
\label{sec:app_A}

%
%
In section \ref{subsec:dist_comp}, 
we gave the solution (\ref{eq:cyl_sol})
of the two dimensional diffusion equation 
for a source of a power-law electron spectrum 
with boundaries defined by parallel planes. 
However, since it has the form of a series, 
it is tedious for numerical evaluation and 
hard to understand the physical meaning of each term. 
We show that we can simplify the situation 
at high energies
by substituting a three dimensional solution without boundaries 
for the two dimensional solution with boundaries. 
In this appendix, 
to see a general behavior of the solution, 
we treat a source of a power-law spectrum without an exponential
cut-off. 
The arguments are valid for $E<<E_{\rm c}$, 
that is $E<{\sim}1$TeV for $E_{\rm c}=10-20$TeV.

Electrons propagate an average distance of 
$R=\sqrt{2DT}$ during their lifetime of $T=1/bE$. 
If the propagation distance is smaller than 
the galactic halo thickness $h$, a
three dimensional solution without boundaries is 
applicable. 
Assuming that a source is located on the plane of the Galactic disk, 
the solution of the diffusion equation (\ref{eq:diffusion_eq}) 
is given by 
\begin{equation}
N_{\rm e}(E,r,t) = \frac{1}{(4{\pi}D_1)^{3/2}}e^{-r^2/(4D_1)}
 Q(\frac{E}{1-bEt})(1-bEt)^{-2}, 
\end{equation}
where 
\[
 D_1 = {\int_{E}^{E/(1-bEt)}}(1/b)DE^{-2}dE. 
\]
Here, assuming that the diffusion coefficient has 
the form $D = D_0E^{\delta}$ 
and a source spectrum of the form $Q(E) = Q_0E^{-\gamma}$, 
we have (e.g. \citealt{ginzburg76} and other references therein) 
\begin{equation}
N_{\rm e}(E,r,t) = \frac{Q_0}{(4{\pi}D_1)^{3/2}}e^{-r^2/(4D_1)}
 (1-bEt)^{\gamma-2} E^{-\gamma}, 
 \label{eq:sph_sol}
\end{equation}
where 
\[
 D_1 = \frac{D_0(1-(1-bEt)^{1-\delta})}{b(1-\delta)E^{1-\delta}}.
\]

Integrating the solution $N_{e}(E,r,t)$ (\ref{eq:sph_sol}) 
on $r$ from 0 to $\infty$ and on $t$ from 0 to $1/bE$, 
the electron flux $J_3$ (from sources assumed to be continuously and uniformly 
distributed in the Galactic disk) is given by 
\begin{equation}
J_3(E) = \frac{c}{4\pi} fQ_{0} 
 \frac{E^{-\gamma-(1+\delta)/2}}{(4{\pi}D_{0}(1-\delta)b)^{1/2}}
 B[(\gamma-1)/(1-\delta),1/2], 
 \label{eq:sol_J3}
\end{equation}
where $B$ is the Beta function.

We compared this three dimensional solution of $J_3$ 
with the two dimensional solution of $J$
with boundaries described in formula (\ref{eq:cyl_sol}). 
Table \ref{tab:J3toJ} shows ratios of $J_3$ to $J$ 
for halo thicknesses of 1kpc to 5kpc and 
electron energies of 1GeV to 10TeV,  
taking the case of a diffusion coefficient of 
$D=2{\times}10^{29}(E/{\rm TeV})^{0.3} {\rm cm}^2{\rm s}^{-1}$ 
as an example in the TeV energy region 
with formula (\ref{eq:lowD}) below 50GeV. 
For larger halo thicknesses or electron energies, 
the differences between $J_3$ and $J$ become smaller. 
For the case where halo thickness $h=3$kpc, 
it is shown that 
the two solutions above 10GeV region agree 
with each other within an error of 1\%.

\begin{center}
\vspace*{0.5cm}
\framebox[3cm]{Table \ref{tab:J3toJ}}
\vspace*{0.5cm}
\end{center}

\section{Evaluation of Cosmic-ray Electron Flux from Distant Sources}

We define the total flux of electrons as $J(E)$,
the flux from distant source at large distances or relatively old ages 
as $J_{\rm d}$, 
and the flux from nearby sources as $J_{\rm n}$. 

To avoid statistical fluctuations in the flux due to the small number of 
nearby sources,
we evaluate $J_{\rm d}$ as shown in section \ref{subsec:dist_comp} 
in the following, 
\begin{equation}
J_{\rm d}(E) = J(E) - J_{\rm n}(E),
\end{equation}
where
\begin{equation}
J_{\rm n} = \frac{c}{4{\pi}} \int _{0}^{T_0} dt 
 \int _{0}^{R_0} dr fN_{\rm e}(E,r,t)2{\pi}r
\end{equation}
is formula (\ref{eq:Jlocal}) defined 
in section\ref{subsec:dist_comp}.

Then, we have
\begin{eqnarray}
\label{eq:domains}
J_{\rm d}(E) &=& \frac{c}{4{\pi}} \{
\int_{0}^{\infty} dt \int _{0}^{\infty} dr fN_{\rm e}(E,r,t)2{\pi}r 
- \int _{0}^{T_0} dt \int _{0}^{R_0} dr fN_{\rm e}(E,r,t)2{\pi}r \}
\nonumber \\ 
             &=& \frac{c}{4{\pi}} \{
\int_{T_0}^{\infty} dt \int_{0}^{\infty} dr fN_{\rm e}(E,r,t)2{\pi}r 
 + \int_{0}^{T_0} dt \int_{R_0}^{\infty} dr fN_{\rm e}(E,r,t)2{\pi}r \} 
\end{eqnarray} 
The integrations are carried out in the domain 1 and 2,
where 
domain 1 is
$T_0<t<{\infty}$ and $0<r<{\infty}$, 
and 
domain 2 is
$0<t<T_0$ and $R_0<r<{\infty}$, respectively. 
In the following, we discuss how to find the appropriate values of 
$R_0$ and $T_0$ for separation of distant and nearby source components.

Taking the three dimensional solution of the formula (\ref{eq:sph_sol}),
the total flux $J(E)$ is given by the formula (\ref{eq:sol_J3}) as 
\begin{equation}
J(E) = \frac{c}{4\pi} fQ_{0} 
 \frac{E^{-\gamma-(1+\delta)/2}}{(4{\pi}D_{0}(1-\delta)b)^{1/2}}
 B[(\gamma-1)/(1-\delta),1/2], 
\label{eq:sol_J3app}
\end{equation}
in the case of a power-law source spectrum, 
where $B$ is the Beta function. 
Then the energy spectrum extends with a power-law 
of the form $E^{-\gamma-(1+\delta)/2}$ without cutoff, 
as is already known in the literatures (e.g. \citealt{berezinskii90}).

The first term in formula (\ref{eq:domains}) (domain 1) 
is given by 
\begin{equation}
\frac{c}{4{\pi}} fQ_0 
\frac{ E^{-{\gamma}-(1+{\delta})/2} }{ (4{\pi}D_0(1-{\delta})b)^{1/2} }
B[(1-bET_0)^{1-{\delta}}, \frac{{\gamma}-1}{1-\delta},\frac{1}{2}], 
\label{eq:dom1term2}
\end{equation}
where $B$ is the incomplete Beta function.
The fraction of the flux from domain 1 to the total flux, 
i.e. (\ref{eq:dom1term2})/(\ref{eq:sol_J3app}), 
should be small enough for the justification on the separation 
of nearby sources from distant sources. 
Numerical value of the fraction
(\ref{eq:dom1term2})/(\ref{eq:sol_J3app}) 
is less than a few percent when $T_{0}=10^{5}$yr and $E>2$TeV.

The contribution from the domain 2 
(the second term in formula (\ref{eq:domains})) is given by 
\begin{equation}
\frac{c}{4{\pi}} fQ_0 
\int_{0}^{T_0} dt (1-bEt)^{{\gamma}-2}
\frac{ E^{-\gamma} }{ ( 4{\pi}D_1 )^{1/2} }
e^{ -\frac{R_0^2}{4D_1} } . 
\label{eq:dom2term3}
\end{equation}
Replacing the variable $t$ with $x$, as
\[
 x = \frac{ 1 }{ 1 - (1-bEt)^{1-{\delta}} },
\]
the integration (\ref{eq:dom2term3}) is reduced to be
\begin{equation}
\frac{c}{4{\pi}} fQ_0
\frac{ E^{-{\gamma}-(1+{\delta})/2} }{ (4{\pi}D_0(1-{\delta})b)^{1/2} }
\int_{ \frac{ 1 }{ 1-(1-bET_0)^{1-{\delta}} } }^{\infty} dx 
(x-1)^{ \frac{{\gamma}-2+{\delta}}{1-{\delta}} } 
x^{ -1.5-\frac{{\gamma}-2+{\delta}}{1-{\delta}} } e^{-x_0 x}, 
\label{eq:dom2term3r}
\end{equation}
where
\[
 x_0 = \frac{ R_0^2 bE(1-{\delta}) }{ 4 D_{0} E^{\delta} } 
   = \frac{1}{2}(\frac{R_0}{R_{\rm a}})^2,  
\]
with 
\[
 R_{\rm a} = (2D_{0}E^{\delta}/(bE(1-{\delta})))^{1/2}. 
\]
The physical meaning of $R_{\rm a}$ is that it represents 
almost the average distance 
traveled by diffusion of electrons with energy $E$.

The upper limit of the integration in formula (\ref{eq:dom2term3r}) 
is obtained by putting the lower boundary of $x$ to be 1.0, i.e. $T_0 = 1/bE$. 
The integral of (\ref{eq:dom2term3r}) 
is then given by the Confluent Hyper-geometric functions, as 
\begin{equation}
\Gamma[\frac{\gamma-1}{1-\delta}]
( \frac{ {\sqrt{\pi}} 
{_{1}F_{1}}[ -0.5-\frac{\gamma-2+\delta}{1-\delta}, 0.5, -x_0 ] }
{ \Gamma[1.5+\frac{\gamma-2+\delta}{1-\delta} ] } 
- \frac{ 2 {\sqrt{\pi}} {x_0^{0.5}} 
{_{1}F_{1}}[ -\frac{\gamma-2+\delta}{1-\delta}, 1.5 , -x_0 ] }
 { \Gamma[\frac{\gamma-1}{1-\delta}] } ), 
\end{equation}
or in another form of the Confluent Hyper-geometric function $U$, as 
\begin{equation}
e^{-x_0} \Gamma[\frac{\gamma-1}{1-\delta}] 
U[\frac{\gamma-1}{1-\delta}, 0.5, x_0]. 
\end{equation}

If the numerical value of $({\gamma}-2+{\delta})/(1-{\delta})$ 
is an integer, 
the integration in formula (\ref{eq:dom2term3r}) 
can be performed without imposing any conditions on $T_0$. 
For example, if we take ${\gamma}=2.4$ and ${\delta}=0.3 $, 
we have $({\gamma}-2+{\delta})/(1-{\delta}) = 1.0$. 
Then the integration of (\ref{eq:dom2term3r}) is carried out, 
and we have 
\begin{eqnarray}
\nonumber
\frac{c}{4{\pi}} fQ_0 \frac{ E^{-3.05} }{ (4{\pi}{D_0}(0.7)b)^{1/2} }
 &(& x_0^{0.5}  {\Gamma}[-0.5, \frac{ x_0 }{ 1-(1-bET_0)^{0.7} }] \\
 &-& x_0^{1.5}  {\Gamma}[-1.5, \frac{ x_0 }{ 1-(1-bET_0)^{0.7} }] ), 
\label{eq:dom2term3rr}
\end{eqnarray}
where $\Gamma$ is the incomplete Gamma function.

Here we define the ratio $F$ 
of the flux in domain 2 to the total flux as 
\begin{equation}
 F = \frac{3}{4} ( 
  x_0^{0.5}  {\Gamma}[-0.5, \frac{ x_0 }{ 1-(1-bET_0)^{0.7} }]
  - x_0^{1.5}  {\Gamma}[-1.5, \frac{ x_0 }{ 1-(1-bET_0)^{0.7} }] ), 
\end{equation}
where we used $B[2,1/2]=4/3$ for the total flux 
from formula (\ref{eq:sol_J3app}). 
$F$ is a function of $x_0$, and gives the relative contribution 
to the total flux 
for a given value of $bET_0$ from sources in domain 2.
The numerical values of $F$ is listed in Table \ref{tab:dom2contrib}.
As shown in Table \ref{tab:dom2contrib}, 
we need to take $x_0>0.8$ to reduce the contribution 
from sources in domain 2 to less than 6\% of the total flux. 
This means we need to take the boundaries of domain 2 as 
$R_0 > \sqrt{2}R_{\rm a}$. 
The average travel distance $R_{\rm a}$ is evaluated 
as $0.5-0.8$kpc at $E=2$TeV depending on the values of 
$D_{0}$ here we adopted. 
Thus if we put 1kpc for $R_0$, 
the contribution from domain 2 gives less than 6\% for $E>2$TeV.

In summary, 
the treatment of separation of nearby and distant sources 
is justified in the energy region beyond ${\sim}2$TeV 
if we put $R_0$=1kpc and $T_{0}=10^{5}$yr.

\begin{center}
\vspace*{0.5cm}
\framebox[3cm]{Table \ref{tab:dom2contrib} }
\vspace*{0.5cm}
\end{center}

\clearpage


\clearpage 


\clearpage

\clearpage

\clearpage

\clearpage

\clearpage

\begin{deluxetable}{lllll}
\tablecaption{\label{tab:nearbysnrs} List of nearby SNRs.}
\tablewidth{0pt}
\tablehead{
\colhead{SNR} & \colhead{Distance(kpc)} & \colhead{Age(yr)} & 
\colhead{$E_{\rm max}$(TeV)\tablenotemark{a}} & \colhead{Reference}
}
\startdata
SN185 & 0.95 & $1.8\times10^3$  & $1.7\times10^2$  & \citep{strom94} \\
S147  & 0.80 & $4.6\times10^3$  & 63   & \citep{braun89} \\
HB 21 & 0.80 & $1.9\times10^4$  & 14   & \citep{tatematsu90}; \\
      &      &                  &      & \citep{leahy96} \\
G65.3+5.7 & 0.80 & $2.0\times10^4$  & 13   & \citep{green88} \\
Cygnus Loop & 0.44 & $2.0\times10^4$ & 13   & \citep{miyata94}; \\
            &      &                 &      & \citep{blair99} \\
Vela  & 0.30 & $1.1\times10^4$ & 25 & \citep{caraveo01} \\
Monogem & 0.30 & $8.6\times10^4$  & 2.8  & \citep{plucinsky96} \\
Loop1 & 0.17 & $2.0\times10^5$  & 1.2  & \citep{egger95} \\
Geminga & 0.4 & $3.4\times10^5$ & 0.67 & \citep{caraveo96} \\
\enddata
\tablenotetext{a}{
Maximum energy limited by the propagation of electrons 
in the case of the prompt release after the explosion.
The delay of the release time gives the larger value.}
\end{deluxetable}

\clearpage

\begin{deluxetable}{cccccc}
\tablecolumns{6}
\tablecaption{\label{tab:J3toJ} Ratios of the three dimensional solution 
without boundaries $J_3$ to the two dimensional solution 
with boundaries $J$ for various energies and halo thicknesses.}
\tablewidth{0pt}
\tablehead{ 
\colhead{}    & \multicolumn{5}{c}{Electron Energy} \\ 
\colhead{}    & \cline{1-5} \\
\colhead{$h$(kpc)} & \colhead{1GeV} & \colhead{10GeV} & 
\colhead{100GeV} & \colhead{1TeV} & \colhead{10TeV}
}
\startdata
1 & 3.35 & 1.69 & 1.08 & 1.00 & 1.00 \\
2 & 1.77 & 1.10 & 1.00 & 1.00 & 1.00 \\
3 & 1.30 & 1.01 & 1.00 & 1.00 & 1.00 \\
4 & 1.12 & 1.00 & 1.00 & 1.00 & 1.00 \\
5 & 1.05 & 1.00 & 1.00 & 1.00 & 1.00 \\
\enddata
\end{deluxetable}

\clearpage

\begin{deluxetable}{crrrrrrrrrr}
\tabletypesize{\small}
\tablecaption{\label{tab:dom2contrib} 
Relative contribution $F$ of the flux at $E=\frac{1}{bT_0}$ 
in domain 2\ ~{\tablenotemark{a}}\ ~to the total flux.}
\tablewidth{0pt}
\tablehead{
\colhead{}    &  \multicolumn{10}{c}{
$x_{0} = \frac{1}{2}(\frac{R_0}{R_{\rm a}})^{2}$\ ~\tablenotemark{b}} \\ 
\colhead{}    & \colhead{0.2} & \colhead{0.4} & \colhead{0.6} & 
\colhead{0.8} & \colhead{1.0} & \colhead{1.2} & \colhead{1.4} & 
\colhead{1.6} & \colhead{1.8} & \colhead{2.0}
}
\startdata
$F$   & 0.2722 & 0.1481 & 0.0901 & 0.0580 & 0.0387 & 0.0265 & 0.0185 & 
0.0131 & 0.0094 & 0.0068 \\
\enddata
\tablenotetext{a}{ $0<t<T_{0}$, $R_{0}<r<{\infty}$.}
\tablenotetext{b}{ See text for the definition of the domain 
and the parameter $x_0$.}
\end{deluxetable}




\end{document}